\documentclass[%
 reprint,
 amsmath,amssymb,
 aps,
]{revtex4-1}

\usepackage{graphicx}% Include figure files
\usepackage{dcolumn}% Align table columns on decimal point
\usepackage{bm}% bold math
\usepackage{natbib}
\begin{document}

\preprint{APS/123-QED}

\title{Accretion disks around black holes in  Scalar-Tensor-Vector Gravity}% Force line breaks with \\

\author{Daniela P\'erez}
\email{danielaperez@iar.unlp.edu.ar}
 \author{Federico G. Lopez Armengol}%
 \email{flopezar@iar.unlp.edu.ar}
 \author{Gustavo E. Romero}
 \email{romero@iar.unlp.edu.ar}
\altaffiliation[Also at ]{Facultad de Ciencias Astron\'omicas y Geof{\'\i}sicas, UNLP, Paseo del Bosque s/n CP, 1900 La Plata, Buenos Aires, Argentina}%Lines break automatically or can be forced with \\

\affiliation{%
 Instituto Argentino de Radioastronom{\'\i}a \\
 (CCT - La Plata, CONICET; CICPBA), Camino Gral Belgrano Km 40 C.C.5, 1894 Villa Elisa, Buenos Aires, Argentina\\
 }%


%\affiliation{
 %Instituto Argentino de Radioastronom{\'\i}a (CCT - La Plata, CONICET; CICPBA), Camino Gral Belgrano Km 40 C.C.5, 1894 Villa Elisa, Buenos Aires, Argentina
% with \\
%}%
%\affiliation{
%\affiliation{%


\date{\today}% It is always \today, today,

\begin{abstract}
 Scalar Tensor Vector Gravity (STVG) is an alternative theory of gravitation that has successfully explained the rotation curves of nearby galaxies, the dynamics of galactic clusters, and cosmological data without dark matter, but has hardly been tested in the strong gravity regime. In this work, we aim at building radiative models of thin accretion disks for both Schwarzschild and Kerr black holes in STVG theory. In particular, we study stable circular equatorial orbits around stellar and supermassive black holes in Schwarzschild and Kerr STVG spacetimes. We also calculate the temperature and luminosity distributions of accretion disks around these objects.
We find that accretion disks in STVG around stellar and supermassive black holes are colder and less luminous than in GR. The spectral energy distributions obtained do not contradict current astronomical observations.
 \end{abstract}
%\begin{description}
%\end{description}
%\end{abstract}

\pacs{Valid PACS appear here}% PACS, the Physics and Astronomy
\maketitle

%\tableofcontents

\section{Introduction}

General Relativity (GR) is a theory about the interaction of spacetime and other material objects. One hundred years after the discovery of the field equations \cite{ein15}, GR continues proving its extraordinary predictive power \cite{abb+16}. Still, GR is a defective theory as pointed out by Einstein himself with regard to the problem of spacetime singularities \cite{rom13}.  GR also fails at reproducing rotation curves of nearby galaxies, mass profiles of galaxy clusters, some gravitational lensing effects, and cosmological data. In order to accommodate GR to astronomical observations, the theory is modified adding a term with a cosmological constant to the field equations, and the existence of dark matter is postulated. However,  every experiment aimed at directly measuring the properties of such matter has failed in its quest so far \cite{apr+12, lux+13, agn+14}. 

A different approach to explain the astronomical data is to introduce a more drastic modification to the theory of gravitation. Following this path, Moffat\cite{mof06} postulated the Scalar-Tensor-Vector Gravity theory (STVG), also called MOdified Gravity theory (MOG). In STVG, in addtion to the tensor metric field, a vector field is introduced, and the universal gravitational constant $G$, along with the mass $\mu$ of the vector field are considered as dynamical scalar fields. In particular, for weak gravitation, STVG leads to a modified acceleration law that has two key features: first, an enhanced Newtonian acceleration law, quantified by $G = G_{\mathrm{N}} (1+\alpha)$ that has successfully explained the rotation curves of many galaxies \cite{bro+06a, mof+13,mof+15}, the dynamics of galactic clusters \cite{bro+06b, bro+07, mof+14}, and cosmological observation \cite{mof+07}, without dark matter. Second, a repulsive Yukawa force term that counteracts the augmented Newtonian acceleration law at certain scales, in such a way that the Solar System results of GR are recovered.
 
 The strong gravity regime of STVG has just started to be explored \footnote{Recently,  Lopez Armengol and Romero \cite{lop+17} constructed neutron star models for four different equations of state, and were able to put restrictive upper limits on the parameter $\alpha$.}. Moffat \cite{mof15a} found solutions of STVG field equations that represent a static, spherically symmetric black hole and also a stationary, axially symmetric black hole, hereafter called Schwarzschild STVG and Kerr-STVG black holes, respectively. The equations to calculate the black hole shadow and lensing effects for Schwarzschild and Kerr STVG black holes were derived in \cite{mof15a, mof15b}. Hussain and Jamil \cite{hus+15} analysed the dynamics of neutral and charged particles around a Schwarzschild STVG black hole inmersed in a weak magnetic field, and studied the stability of the circular orbits. Recently, John \cite{joh16} studied the spherically symmetric accretion of a gas with a polytropic equation of state into a Schwarzschild STVG black hole. Analytical expressions were obtained for the mass accretion rate, critical velocity, and critical sound speed. 
 
In the present work, we construct thin relativistic accretion disks around Schwarzschild and Kerr black holes in STVG theory. In particular, we calculate temperature and spectral energy distributions of accretion disks around stellar and supermassive black holes, and compare the results with those obtained in GR. We analyse the viability of the theory in the strong gravity regime by contrasting our predictions with current astronomical data.

The paper is organised as follows. In Sect. \ref{sec2} we provide a brief review of STVG theory. We study circular orbits in both Schwarzschild and Kerr STVG spacetimes in Sect. \ref{cine}. Section \ref{sec4} is devoted to the calculation of the properties of accretion disks in these spacetimes, and in Sect. \ref{sec5} we analyse the results previously obtained. We offer the conclusions of the work in the last section.

%______________________________________________________________________________________________________________________________________________________

\section{\label{sec2}STVG gravity}

\subsection{STVG action and field equations}

The action \footnote{As suggested in \citet{mof+13} and \cite{mof+09}, we dismiss the scalar field $\omega$, and we treat it as a constant, $\omega =1$.} in STVG theory is \cite{mof06}:
\begin{equation}\label{action}
S = S_{\rm GR} + S_{\phi} + S_{\rm S} + S_{\rm M},
\end{equation}
where
\begin{eqnarray}
S_{\rm GR} & = & \frac{1}{16 \pi} \int d^{4}x \sqrt{-g} \frac{1}{G} R ,\\
S_{\phi} & = & - \int d^{4}x \sqrt{-g}  \left(\frac{1}{4} B^{\mu\nu}B_{\mu\nu} - \frac{1}{2}\tilde{\mu}^2 \phi^{\mu} \phi_{\mu}\right),\\
S_{\rm S} & = & \int d^{4}x \sqrt{-g} \frac{1}{G^{3}} \left(\frac{1}{2} g^{\mu \nu} \nabla_{\mu}G\nabla_{\nu}G-V(G)\right)\\
%& + & \int d^{4}x \frac{1}{G} \left(\frac{1}{2} g^{\mu \nu} \nabla_{\mu}\omega\nabla_{\nu}\omega-V(\omega)\right)\\
& + &  \int d^{4}x \frac{1}{\tilde{\mu}^2 G} \left(\frac{1}{2} g^{\mu \nu} \nabla_{\mu}\tilde{\mu}\nabla_{\nu}\tilde{\mu}-V(\tilde{\mu})\right).
\end{eqnarray}
Here, $g_{\mu \nu}$ is the spacetime metric, $R$ denotes the Ricci scalar, and $\nabla_{\mu}$ is the covariant derivative; $\phi^{\mu}$ stands for a Proca-type massive vector field, $\tilde{\mu}$ is
its mass, and $B_{\mu \nu} = \partial_{\mu} \phi_{\nu} - \partial_{\nu}\phi_{\mu}$; $G(x)$,  
and $\tilde{\mu}(x)$ are scalar fields that vary in space and time, and $V(G)$, and $V(\tilde{\mu})$ 
are the corresponding potentials. We adopt the metric signature $\eta_{\mu \nu} = {\rm diag}(-1,+1,+1,+1)$. 
The term $S_{\rm M}$ in the action refers to possible matter sources.

The full energy-momentum tensor for the gravitational sources is:
\begin{equation}
T_{\mu \nu} = T^{\rm M}_{\mu \nu} + T^{\phi}_{\mu \nu} + T^{\rm S}_{\mu \nu},
\end{equation} 
where
\begin{eqnarray}
T^{\rm M}_{\mu \nu} & = & -\frac{2}{\sqrt{-g}} \frac{\delta S_{\rm M}}{\delta g^{\mu \nu}},\\
T^{\phi}_{\mu \nu} & = & -\frac{2}{\sqrt{-g}} \frac{\delta S_{\phi}}{\delta g^{\mu \nu}},\\
T^{\rm S}_{\mu \nu} & = & -\frac{2}{\sqrt{-g}} \frac{\delta S_{\rm S}}{\delta g^{\mu \nu}}.
\end{eqnarray}
Following the notation introduced above, $T^{\rm M}_{\mu \nu}$ denotes the ordinary matter energy-momentum 
tensor and $T^{\phi}_{\mu \nu}$ the energy-momentum tensor of the field $\phi^{\mu}$; $T^{\rm S}_{\mu \nu}$
 denotes the scalar contributions to the energy-momentum tensor.

The Schwarzschild and Kerr STVG black hole solutions were found under the following assumptions \cite{mof15a}:
\begin{itemize}
%\item The dimensionless scalar field $\omega$ is treated as a constant, and it is set $\omega = 1$ \citep{mof+13}.\\
\item The mass of the vector field $\phi^{\mu}$ is neglected; since the effects of $m_{\phi}$ manifest at kiloparsec scales from the source\footnote{The mass of the field $\phi^{\mu}$ determined from galaxy rotation curves and galactic cluster dynamics is $\tilde{\mu} = 0.042$ kpc$^{-1}$
 \cite{mof+13,mof+14,mof+15}.}, it can be dismissed when solving the field equations for compact objects such as black holes.\\
\item $G$ is a constant that depends on the parameter $\alpha$ \cite{mof06}:
\begin{equation}
G = G_{\rm N} \left(1 + \alpha \right),
\end{equation} 
where $G_{\rm N}$ denotes Newton's gravitational constant, and $\alpha$ is a free adimensional parameter.
\end{itemize}

Given these hypotheses, the action \eqref{action} takes the form,
\begin{equation}\label{action1}
S =   \int d^{4}x \sqrt{-g} \left( \frac{R}{16 \pi G} -\frac{1}{4}B^{\mu\nu} B_{\mu\nu}\right).
\end{equation}
Variation of the latter expression with respect to $g_{\mu \nu}$ yields the STVG field equations:
\begin{equation}
G_{\mu \nu} = 8 \pi G T^{\phi}_{\mu \nu},
\end{equation}
where $G_{\mu \nu}$ is the Einstein tensor, and the energy-momentum tensor for the vector field $\phi^{\mu}$ is given by\footnote{ Moffat \cite{mof15a} set the potential $V(\phi)$ equal to zero in the definition of $T^{\phi}_{\mu \nu}$ given in \cite{mof06}.}
\begin{equation}
T^{\phi}_{\mu \nu} = - \frac{1}{4}\left({B_{\mu}}33^{\alpha} B_{\nu \alpha} - \frac{1}{4} g_{\mu \nu} B^{\alpha \beta} B_{\alpha \beta}\right).
\end{equation}
If we vary the action \eqref{action1} with respect to the vector field $\phi^{\mu}$, we obtain the dynamical equation for such field:
\begin{equation}\label{tb}
\nabla_{\nu}B^{\mu \nu} = 0.
\end{equation}
%whose vector potential in Boyer-Lindquist coordinates is:
%\begin{equation}
%\mathbf{A} = \frac{- Q r}{\rho^{2}} \left( \mathbf{dt} - a \sin^{2}{\theta} \ \mathbf{d\phi} \right).  
%\end{equation}

The equation of motion for a test particle in coordinates $x^{\mu}$ is given by
\begin{equation}\label{eq-motion}
\left(\frac{d^{2}x^{\mu}}{d\tau^2} + \Gamma^{\mu}_{\alpha \beta} \frac{dx^{\alpha}}{d\tau}\frac{dx^{\beta}}{d\tau} \right) =\frac{q}{m} {B^{\mu}}_{\nu} \frac{dx^{\nu}}{d\tau},
\end{equation}
where $\tau$ denotes the particle proper time, and $q$ is the coupling constant with the vector field.

Moffat \cite{mof15a}  postulates that the gravitational source charge $q$ of the vector field $\phi^{\mu}$ is proportional to the mass of the source particle,  
\begin{equation}
q = \pm \sqrt{\alpha G_{N}} m.
\end{equation}
The positive value for the root is chosen ($q > 0$) to maintain a repulsive, gravitational Yukawa-like force when the mass parameter $\tilde{\mu}$ is non-zero (see the Appendix for furhter details). Then, we see that in STVG theory the nature of the gravitational field has been modified with respect to GR in two ways: an enhanced gravitational constant $G = G_{\rm N} \left(1 + \alpha \right)$, and a vector field $\phi^{\mu}$ that exerts a gravitational Lorentz-type force on any material object through Eq. \eqref{eq-motion}.

% Notice, that there is no electric charge in STVG theory.%necessary to describe physically stable stars, galaxies, galaxy clusters and agreement with solar system observational data \cite{Moffat-BH}.

In what follows, we present the Schwarzschild and Kerr STVG black holes and analyse their main properties.

%_______________________________________________________________________________________________________________________________________________

\subsection{STVG black holes}
\subsubsection{Schwarzschild STVG black hole}   

The line element of the spacetime metric for a Schwarzschild STVG black hole is given by \cite{mof15a}:
\begin{eqnarray}
ds^{2} & = & -  \left(1 -  \frac{ 2 G_{N} (1+\alpha) M}{c^{2} r} + \frac{G_{N}^{2} M^{2} (1+\alpha) \alpha}{c^{4} r^{2}}\right) dt^{2}\nonumber \\ 
& +&   \left(1 - \frac{ 2 G_{N} (1+\alpha) M}{c^{2} r} + \frac{G_{N}^{2} M^{2} (1+\alpha) \alpha}{c^{4} r^{2}}\right)^{-1} dr^2 \nonumber \\
&+  & r^2 \left(d\theta^{2} + \sin{\theta}^{2} d\phi^{2}\right),\label{schw}
\end{eqnarray}
where $M$ stands for the mass of the black hole, and $\alpha$ determines the strength of the gravitational vector forces. For $\alpha \rightarrow 0$, the Schwarzschild metric in GR is recovered. 

The metric \eqref{schw} has an outer ($r_{+}$) and an inner ($r_{-}$) event horizons:
\begin{equation}\label{hori-schw}
r_{\pm}  = \frac{G_{N} M}{c^2} \left[1 + \alpha \pm \left(1+\alpha\right)^{1/2}\right].
\end{equation}
If we set $\alpha = 0$, then $r_{+} = r_{s} = 2GM/c^2$ is the Schwarzschild event horizon in GR. From Eq. \eqref{hori-schw} we see that if $\alpha > 0$
the radius of the Schwarzschild STVG black hole outer horizon is larger than the corresponding horizon in GR. 

Next, we introduce the spacetime metric for the Kerr STVG black hole.

\subsubsection{Kerr STVG black hole}

The line element of the spacetime metric of a black hole of mass $M$ and angular momentum $J = a M$ in STVG theory is \cite{mof15a}:
\begin{eqnarray}\label{BH-metric}
ds^{2} & = &  -c^{2} \left(\Delta - a^{2} \sin^{2}\theta\right) \frac{dt^{2}}{\rho^{2}} + \frac{\rho^{2}}{\Delta} dr^{2} + \rho^{2} d\theta^{2}\nonumber \\
& + &  \frac{2 a c \sin^{2}{\theta}}{\rho^{2}}\left[(r^{2}+a^{2})-\Delta\right] dt d\phi \nonumber \\
&+ & \left[(r^{2}+a^{2})^{2}-\Delta a^{2} \sin^{2}\theta\right] \frac{\sin^{2}\theta}{\rho^{2}} d\phi^{2},\\
\Delta & = & r^{2} - \frac{2 G M r}{c^{2}}+ a^{2} + \frac{\alpha G_{N} G M^{2}}{c^{4}}\label{Delta} \\
& = & r^{2} - \frac{2 G_{N} (1+\alpha) M r}{c^{2}}+ a^{2} + \frac{\alpha (1+\alpha) G_{N}^{2} M^{2}}{c^{4}},\nonumber\\
\rho^{2} & = & r^{2} + a^{2} \cos^{2}\theta.
\end{eqnarray}
The metric above reduces to the Kerr metric in GR when $\alpha = 0$. By setting $a = 0$ in Eq. \eqref{BH-metric}, we recover the metric of a Schwarzschild STVG black hole. 

The radius of the inner $r_{-}$ and outer $r_{+}$ event horizons are determined by the roots of $\Delta = 0$:
\begin{equation}\label{hori-Kerr}
r_{\pm} = \frac{G_{N} (1+\alpha)M}{c^{2}}\left\{1 \pm \sqrt{1- \frac{a^{2} c^{4}}{G_{N}^{2} (1+\alpha) M^{2}} - \frac{\alpha}{(1+\alpha)}}\right\}. 
\end{equation}
If we set $\alpha =  0$ in the latter expression, we obtain the formula for the inner and outer horizons of a Kerr black hole. Inspection of Eq. \eqref{hori-Kerr} also reveals that for $\alpha > 0$ the outer horizon of a Kerr black hole in STVG is larger that the corresponding one in GR.

The radial coordinate of the ergosphere is determined by the roots of $g_{tt} =0$:
\begin{equation}
r_{E} = \frac{G_{\rm N} M \left(1+\alpha\right)}{c^{2}}\left[1 \pm \sqrt{1 - \frac{a^{2} \cos^{2}{\theta} \ c^{4}}{{G_{N}}^{2} \left(1+\alpha\right)^{2} M^{2} }-\frac{\alpha}{1+\alpha}}\right].  
\end{equation}
We see that the ergosphere grows in size as the parameter $\alpha$ increases. 

In the next section, we calculate and analyse the equatorial circular motion of test particles around these black holes. We develop the formalism for the calculation of the circular orbits in Kerr STVG spacetime. The corresponding expressions in Schwarzschild STVG spacetime can be easily obtained by setting $a =0$.

%________________________________________________________________________________________________________________________________________

\section{\label{cine}Circular orbits around STVG black holes} 
   
The radial motion of a test particle of mass $m$ in the equatorial plane $\theta = \pi/2$ of a Kerr STVG black hole satisfies the energy equation\footnote{Since the Kerr STVG metric formally resembles the Kerr-Newman metric in GR, the calculation of the trajectories of a test particle in Kerr STVG spacetime is mathematically equivalent to the computation of the trajectories of a test particle of electric charge $q$ in Kerr-Newman spacetime in GR. In this section, we adapt the formalism given in \cite{mis+13} to STVG.}:
\begin{equation}\label{energy1}
\chi(r,\alpha) E^{2} -2 \beta(r,L,\alpha) E + \gamma_{0}(r,L,\alpha) + r^{4} \left({p^{r}}\right)^{2} = 0,
\end{equation}
where $E$ is the energy of the particle, $p^{r} = m \ dr/ds$ the radial momentum, and:
\begin{eqnarray}
\chi(r,\alpha) & = & \left(r^{2}+a^{2}\right)^{2}-\Delta a^{2}, \label{chi} \\
\beta(r,L,\alpha) & = & \left(L \:a + \frac{\alpha G_{\rm N}\: m \: M \: r}{c}\right) \left(r^{2}+a^{2}\right)\nonumber \\
& -& L\: a \Delta,\label{beta}\\
\gamma_{0}(r, L,\alpha) & = & \left(L \: a + \frac{\alpha G_{\rm N} \: m \: M \: r}{c}\right)^{2}- \Delta \: L^{2}\nonumber\\ 
& - & c^{2} r^{2} m^{2} \Delta. \label{gamma} 
\end{eqnarray} 
In the above formulas, $L$ denotes the angular momentum of the test particle. The solution of Eq. \eqref{energy1} is:
\begin{equation}\label{energy2}
E = \frac{\beta(r,L,\alpha) + \sqrt{\beta(r,L,\alpha)^{2} - \chi(r,\alpha) \gamma_{0}(r, L,\alpha) - r^{4} {p^{r}}^{2}}}{\chi(r,l,\alpha)}.
\end{equation}
Since we aim at studying circular orbits, we set $p^{r} =0$ in the latter equation, and we derive an expression for the effective potential $V(r, L, \alpha)$:
\begin{equation}
V(r,L,\alpha) = \frac{\beta(r,L,\alpha) + \sqrt{\beta(r,L,\alpha)^{2} - \chi(r,\alpha) \gamma_{0}(r, L,\alpha)}}{\chi(r,L,\alpha)}.
\end{equation}
The allowed regions for the motion of a test particle with energy $E$ at infinity are the regions with $V(r) \leq E$, and the turning points ($p^{r} = dr/ds = 0$) occur for $V(r) = E$.

We compute the radius of the circular orbits by solving the equation:
\begin{equation}\label{derivada-pot}
\frac{dV(r,L,\alpha)}{dr} = 0,
\end{equation}
for a given value of $L$, and $\alpha$.

We express the effective potential in terms of the following dimensionless quantities:
\begin{eqnarray}
%r_{g} & = & \frac{G_{N}M}{c^{2}},\\
%\tilde{r_{g}} & = & \frac{G M}{c^{2}} = \frac{G_{N}(1+\alpha)M}{c^{2}} = (1+\alpha) r_{g},\\
x & = & \frac{r}{r_{g}},\\
%\frac{r}{\tilde{r_{g}}} & = & (1+\alpha)^{-1}\frac{r}{r_{g}} = (1+\alpha)^{-1} x,\\
\tilde{a} & = & \frac{a}{r_{g}},\\
\tilde{L} & = & \frac{L}{m c r_{g}}
\end{eqnarray}
where $r_{g} = G_{N} M /c^2$ is the gravitational radius.

Equations \eqref{Delta}, \eqref{chi}, \eqref{beta}, and \eqref{gamma} are now written as:
\begin{eqnarray}
\tilde{\Delta}(x,\alpha) & = & x^{2}-2 x (1+\alpha)+\tilde{a}^{2}+\alpha (1+\alpha), \label{delta}\\
\tilde{\beta}(x,\tilde{L},\alpha) & = &  \left(\tilde{L}\:\tilde{a}+\alpha x\right) (x^2+\tilde{a}^{2})\nonumber\\
& - & \tilde{L} \ \tilde{a} (1+\alpha)^{2}\tilde{\Delta}(x,\alpha), \\
\tilde{\gamma_{0}}(x,\tilde{L},\alpha) & = & \left(\tilde{L}\:\tilde{a}+\alpha x\right)^{2}\nonumber\\
& -& (x^2+\tilde{L}^{2}) (1+\alpha)^{2}\tilde{\Delta}(x,\alpha), \\
\tilde{\chi}(x,\alpha) & = & (x^{2}+a^{2})^{2}- (1+\alpha)^{2} \tilde{\Delta}(x,\alpha) \tilde{a}^{2}.
\end{eqnarray}

We have derived all the necessary formulae to calculate the value of the radial coordinate of the last stable circular orbit around a Kerr STVG black hole of mass $M$, and angular momentum $J = a \ M$. We now need to estimate the range of values of the parameter $\alpha$ that corresponds to stellar and supermassive black holes.

The value of the parameter $\alpha$ depends on the mass of the gravitational central source. Since we aim at modeling accretion disks around stellar and supermassive black holes, we first need to set the range of the permissible values of the parameter $\alpha$.

For stellar mass sources, Lopez and Romero \cite{lop+17} found that:
\begin{equation}\label{alfa-c1}
\boxed{\alpha <  0.1.}
\end{equation}

For supermassive black holes ($10^{7} M_{\odot} \leq M \leq 10^{9} M_{\odot}$), we estimate the lower and upper limit for $\alpha$ as follows:
\begin{itemize}
\item \textit{Lower limit for $\alpha$}: Moffat et al. \cite{mof+08} showed that for globular clusters ($ 10^{4} M_{\odot} \leq M \leq 10^{6} M_{\odot}$) the predicted value of the parameter $\alpha$ is $\alpha = 0.03$.\\

\item \textit{Upper limit for $\alpha$}: Brownstein et al. \cite{bro+06b} placed restrictions on the parameter $\alpha$ by fitting rotation curves of dwarf galaxies ($ 1.9 \times 10^{9} M_{\odot} \leq M \leq 3.4 \times 10^{10} M_{\odot}$). They obtained:
\begin{equation}
2.47 \leq \alpha \leq 11.24.
\end{equation}
\end{itemize}

Since the values of the masses of supermassive black holes lay in between the masses of globular clusters and dwarf galaxies, we assume that for such black holes:
\begin{equation}\label{alfa-c2}
  \boxed{0.03 < \alpha < 2.47.}
\end{equation}

\begin{table*}
\caption{\label{table:1}Values of the relevant parameters. The angular momentum $\tilde{a}$ has been normalised as $\tilde{a}= a / r_{g}$; here, $a$ has unit of length.}             % title of Table
      % is used to refer this table in the text
%\centering     
\begin{ruledtabular}
\begin{tabular}{c c c }        % centered columns (4 columns)
            % inserts double horizontal lines
Parameters, symbol [units]&  Stellar mass BH & Supermassive BH \\    % table heading 
\hline                     % inserts single horizontal line
Mass, $M$ $[M_{\odot}]$ & 14.8 & $10^{8}$  \\      % inserting body of the table
Angular momentum $\tilde{a} $& 0.99 & 0.99     \\
Mass accretion rate, $\dot{M}$ $[\dot{M}_{\mathrm{Edd}}]$ & 2.53 & 0.5     \\
\end{tabular}
\end{ruledtabular}
\end{table*}

Given the conditions \eqref{alfa-c1} and \eqref{alfa-c2}, we now proceed to calculate the location of the innermost stable circular orbit (ISCO) in Schwarzschild and Kerr STVG spacetimes, respectively. The stellar and supermassive black hole are characterised by the parameters displayed in Table \ref{table:1}. For the stellar mass black hole, we adopt the best estimates available for the well-known galactic black hole Cygnus-X1 \cite{oro+11,gou+11}.

\begin{figure}
\center
\resizebox{\hsize}{!}{\includegraphics{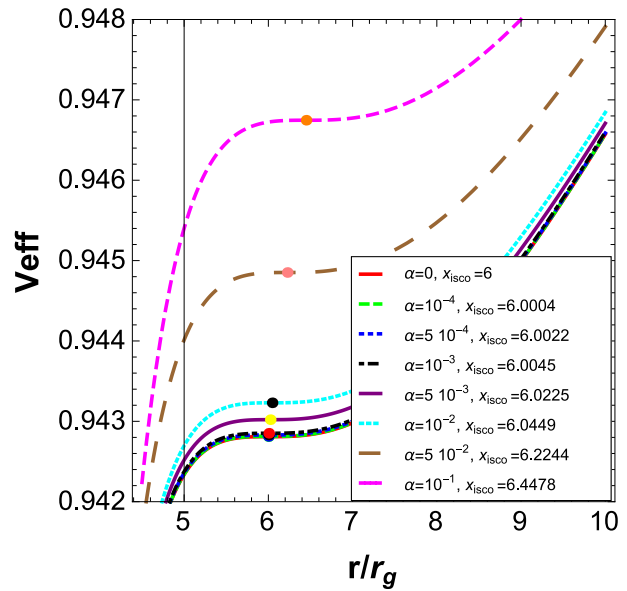}}% Here is how to import EPS art
\caption{Plot of the effective potential as a function of the radial coordinate for different values of the parameter $\alpha$ in the case of a stellar mass black hole in Schwarzschild STVG spacetime. Here, $x_{ {\rm isco}} =  r_{{\rm isco}} / r_{{\rm g}}$ stands for the location of the innermost stable circular orbit.}
         \label{fig1}
   \end{figure}

\begin{figure}
   \center
   \resizebox{\hsize}{!}{\includegraphics{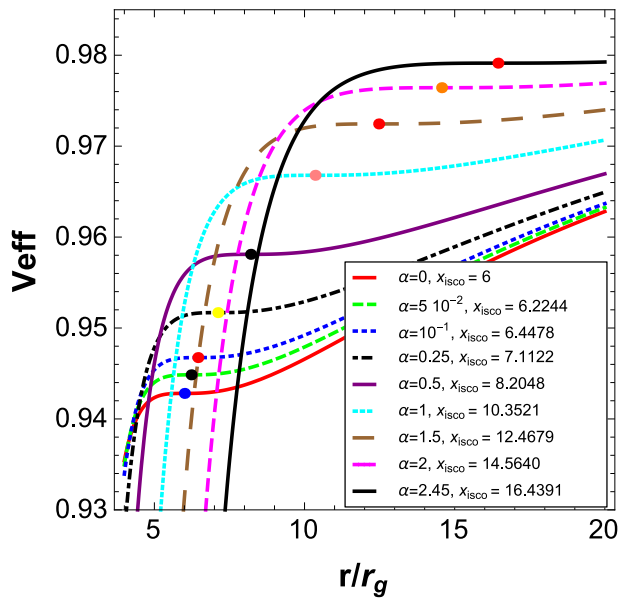}}
      \caption{Plot of the effective potential as a function of the radial coordinate for different values of the parameter $\alpha$ in the case of a supermassive black hole in Schwarzschild STVG spacetime. Here, $x_{{\rm isco}} =  r_{ {\rm isco}} / r_{{\rm g}}$ stands for the location of the innermost stable circular orbit.}
         \label{fig2}
   \end{figure}

In Figs. \ref{fig1} and \ref{fig2}, we show the plot of the effective potential as a function of the radial coordinate in Schwarzschild STVG spacetime for a stellar and supermassive black hole. In both cases, and for all values of the parameter $\alpha$, the ISCO is larger than for a Schwarzschild black hole in GR. This diference, however, is minor for stellar black holes: at most $0.45$ percent. On the contrary, the last stable circular orbit for a supermassive black hole in STVG spacetime can lay as far as $16.44 \ r_{\rm g}$, a departure of $174$ percent with respect to GR.

% \begin{figure}
  % \centering
   %\includegraphics[bb=10 20 100 300,width=5cm,clip]{grafico-pot-eff-stellar.eps}

\begin{figure}
   \center
  \resizebox{\hsize}{!} {\includegraphics{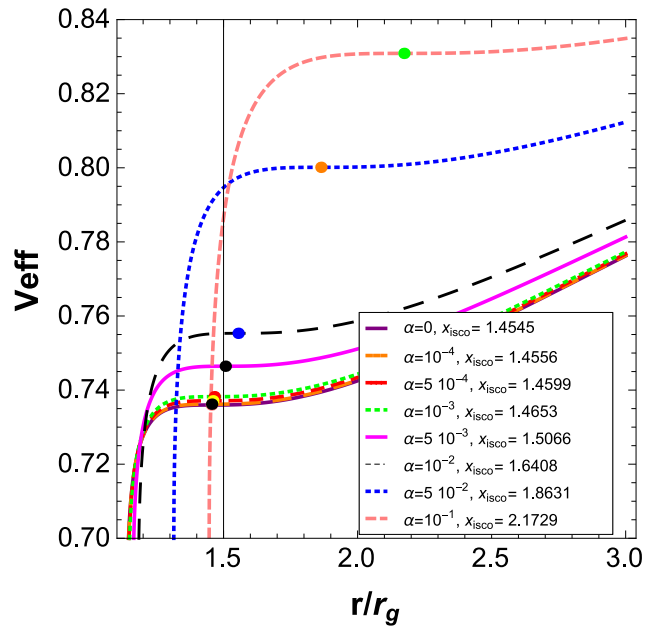}}
      \caption{Plot of the effective potential as a function of the radial coordinate for different values of the parameter $\alpha$ in the case of a stellar mass black hole in Kerr STVG spacetime; the spin of the black hole is $\tilde{a} = 0.99$. Here, $x_{{\rm isco}} =  r_{{\rm isco}} / r_{{\rm g}}$ stands for the location of the innermost stable circular orbit.}
         \label{fig3}
   \end{figure}

\begin{figure}
   \center
 \resizebox{\hsize}{!}{\includegraphics{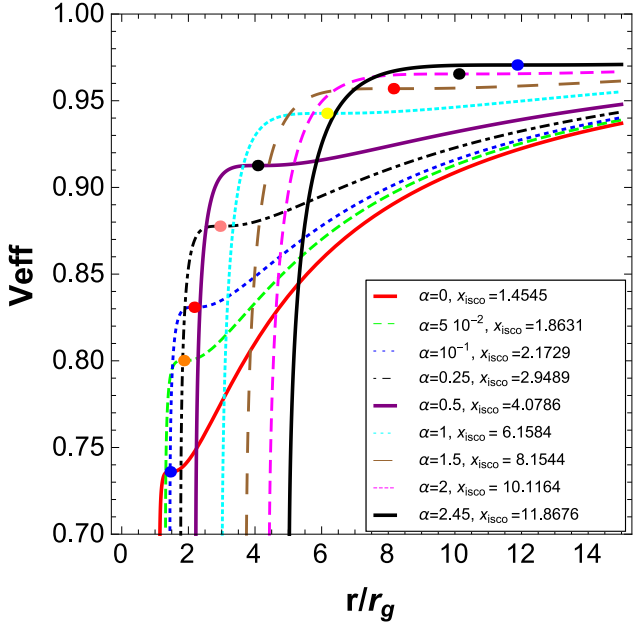}}
      \caption{Plot of the effective potential as a function of the radial coordinate for different values of the parameter $\alpha$ in the case of a supermassive black hole in Kerr STVG spacetime; the spin of the black hole is $\tilde{a} = 0.99$. Here, $x_{{\rm isco}} =  r_{{\rm isco}} / r_{{\rm g}}$ stands for the location of the innermost stable circular orbit.}
         \label{fig4}
   \end{figure}

The last stable circular orbit for Kerr black holes in STVG is also larger that for a Kerr black hole of the same spin in GR (see Figs. \ref{fig3}, and \ref{fig4}). As the value of the parameter $\alpha$ rises, the radius of the ISCO increases. For stellar mass black holes, the radius of the ISCO growths up to $49$ percent with respect to the value of the ISCO in GR; for supermassive black holes, the ISCO increases up to 716 percent for $\alpha = 2.45$. 

%The reason the ISCO in STVG is farther from the black hole than in GR is due to the repulsive character of the vector field $\phi^{\mu}$ and the enhanced gravitational constant $G = G_{N} (1+\alpha)$.

The analysis of the circular orbits presented in this section
will be applied next to the construction of accretion disks around STVG black holes.

%________________________________________________________________________________________________________________________________________________
\section{\label{sec4}Accretion disks in STVG theory}\label{sec4}

\subsection{Analysis of the structure of the disk}

Novikov and Thorne \cite{nov+73} and Page and Thorne \cite{pag+74} made the first relativistic analysis of the structure of an accretion disk around a black hole in General Relativity. They supposed the disk self-gravity is negligible, and that the background spacetime geometry is stationary, axially symmetric, asymptotically flat, and reflection-symmetric with respect to the equatorial plane. They also assumed that the central plane of the disk is located at the equatorial plane of the spacetime geometry; since the disk is supposed to be thin ($\Delta z = 2 h << r$, where $z$ is a coordinate that measures the height above the equatorial plane, and $h$ is the thickness of the disk at radius $r$) the metric coefficients only depend on the radial coordinate $r$.

%The disk is supposed to be in a quasi-steady state \citep{nov+73}, so any relevant quantity (for example the density or the temperature of the gas) is averaged over $2\pi$, a proper radial distance of order $2h$, and the time interval $\Delta t$ that the gas takes to move inward a distance $2h$.

The three fundamental equations for the time-averaged radial structure of the disk were derived in \cite{pag+74}. In particular, an expression for the heat emitted by the accretion disk was obtained that depends on the specific energy $E^{\dag}$, the specific angular momentum $L^{\dag}$, and the angular velocity $\Omega$ of the particles that move on equatorial circular geodesic orbits around the black hole:
\begin{equation}\label{heat-pt}
Q(r) = \frac{\dot{M}}{4\pi \sqrt{-g}}\frac{\Omega_{,r}}{\left(E^{\dag}-\Omega L^{\dag}\right)^{2}} \int^{r}_{r_{\rm isco}}  \left(E^{\dag}-\Omega L^{\dag}\right) {L^{\dag}}_{,r} \ dr, 
\end{equation}
where $\dot{M}$ stands for the mass accretion rate, and $g$ is the metric determinant. The lower limit of this integral corresponds to the radius of the innermost stable circular orbit.

Equation \eqref{heat-pt} is derived from the laws of conservation of rest mass, angular momentum, and energy. The specific quantities are:
\begin{eqnarray}
E^{\dag}(r) & \equiv & - u_{t}(r), \label{ene} \\
L^{\dag} & \equiv & u_{\phi}(r), \label{lz}\\
\Omega(r) & \equiv & \frac{u^{\phi}}{u^{t}},
\end{eqnarray}
where $u_{t}$ and $u_{\phi}$ are the $t$ and $\phi$ components of the four-velocity of the particle, respectively. Expressions \eqref{ene} and \eqref{lz} no longer hold in STVG. Because of the presence of the gravitational vector field $\phi^{\mu}$, a neutral test particle in STVG is subjected to a gravitational Lorentz force, whose vector potential in Boyer-Lindquist coordinates is:
\begin{equation}
\boldsymbol{\phi} = \frac{- Q r}{\rho^{2}} \left( \mathbf{dt} - a \sin^{2}{\theta} \ \mathbf{d\phi} \right).  
\end{equation}
Following \cite{mis+13} we redefine expressions \eqref{ene} and \eqref{lz} as:
\begin{eqnarray}
\tilde{E} & = & - \frac{p_{t}}{m} + \frac{q}{m} \phi_{t},\\
\tilde{L} & = & \frac{p_{\phi}}{m}  +  \frac{q}{m} \phi_{\phi}.
\end{eqnarray}

The conservation laws (Equations 32a, and 32b in \cite{pag+74}) take the form:
\begin{eqnarray}
\left(\tilde{L} - w \right)_{,r} + \frac{q}{m}  {\phi_{\phi}}_{,r} & = & f \left(\tilde{L} + \frac{q}{m}\phi_{\phi} \right),\label{con-l}\\
\left(\tilde{E} - \Omega w \right)_{,r} +  \frac{q}{m} {\phi_{t}}_{,r} & = & f \left(\tilde{E} + \frac{q}{m} \phi_{t}\right),\label{con-e}
\end{eqnarray}
where
\begin{eqnarray}
f & = & 4 \pi e^{\nu+\Phi + \mu} F/{\dot{M}_{0}},\\
w & = & 2 \pi e^{\nu+\Phi + \mu}{W_{\phi}}^{r}/{\dot{M}_{0}}.
\end{eqnarray}
$F$ stands for the emitted flux, ${W_{\phi}}^{r}$ denotes the torque per unit circunference, $e^{\nu+\Phi + \mu} = \left(-g\right)^{1/2}$ is the square root of the metric determinant, and $\dot{M}_{0}$ the mass accretion rate.

If we multiply Eq. \eqref{con-l} by $\Omega$, and substract from Eq. \eqref{con-e}, we obtain the relation:
\begin{equation}\label{w}
w = \frac{\left[\left(\tilde{E}-\Omega \tilde{L}\right)+ q/m \left(\phi_{t} - \Omega \phi_{\phi}\right) -q/m  \left({\phi_{t}}_{,r} - \Omega {\phi_{\phi}}_{,r}\right)  \right]}{-\Omega_{,r}}.  
\end{equation}
The latter expression reduces to Eq. (33) in \cite{pag+74} if $\boldsymbol{\phi} = \mathbf{0}$.

If we insert Eq. \eqref{w} into Eq. \eqref{con-l}, after some algebra, we get a first order differential equation for $f$:
\begin{equation}\label{flux}
a_{0}(x) = a_{1}(x) f + a_{2} (x) \frac{df}{dx},
\end{equation}
where,
\begin{eqnarray}
a_{0}(x) & = & \left(\tilde{E}-  \ \ \tilde{\Omega}\tilde{L}\right)\frac{d}{dx}\left[ \frac{\left(\tilde{E}-  \ \tilde{\Omega}\tilde{L}\right)_{,x}}{-   \tilde{\Omega}_{,x}}\right],\label{a0}\\
a_{1}(x) & = &  \left[\frac{\left(\tilde{E}-  \ \tilde{\Omega}\tilde{L}\right)^2}{-  \ \tilde{\Omega}_{,x}}\right]_{,x}\nonumber \\ 
& -&  \frac{\left(\tilde{E}-  \tilde{\Omega}\tilde{L}\right) \left(\tilde{E}_{,x} - \ \tilde{\Omega} \tilde{L}_{,x}\right)}{- \tilde{\Omega}_{,x}},\label{a1}\\
a_{2}(x) & = & \frac{\left(\tilde{E}-   \tilde{\Omega} \tilde{L}\right)^2}{-  \tilde{\Omega}_{,x}}.\label{a2}
\end{eqnarray}

Below we write the expressions for the specific energy $\tilde{E}$, angular momentum $\tilde{L}$\footnote{We denote by $\tilde{E}$, and $\tilde{L}$ the energy and angular momentum per mass unit, respectively.}, and angular velocity for a massive particle in a circular orbit around a STVG Kerr black hole:
\begin{equation}\label{efinal}
\tilde{E} = - e_{1}\left(e_{2} + \frac{k_{1}}{\sqrt{\beta}}\right),
\end{equation}

\begin{equation}\label{lfinal}
\tilde{L} = - \frac{\tilde{E} \left( \tilde{\Omega} \ \tilde{g_{\phi \phi}} + \tilde{g_{t\phi}}\right)}{\left( \tilde{g_{tt}} +  \tilde{\Omega} \  \tilde{g_{t \phi}}\right)} -  \frac{\alpha  \  \left[ -\left(x^2 + \tilde{a}^2\right) \ \tilde{\Omega}+\tilde{a}\right]}{x \left(  \tilde{g_{tt}} +   \tilde{\Omega} \tilde{g_{t \phi}}\right)},
\end{equation}

where,
\begin{eqnarray}
\beta & = & - g_{tt}  -   \tilde{\Omega} \left( 2 \  \tilde{g_{t \phi}}  +  \  \tilde{g_{\phi \phi}} \tilde{\Omega}\right),\\
k_{1} & = & \tilde{g_{tt}} +  \tilde{\Omega} \tilde{g_{t \phi}},\\
e_{1} & = & - \sqrt{1 + \frac{\left(e_{11}+e_{12}\right) \alpha^{2}}{-x^{2} \left(\tilde{g_{t\phi}}^2 - \tilde{g_{tt}} \tilde{g_{\phi \phi}}\right) }},\\
e_{11} & = & a^{2} \left(a^{2} \tilde{g_{tt}} + 2 a \tilde{g_{t\phi}} + \tilde{g_{\phi \phi}} \right) ,\\
e_{12} & = & x^{2} \left(2 a^{2} \tilde{g_{tt}} + 2 a \tilde{g_{t\phi}} + \tilde{g_{\phi \phi}}^{2} - \tilde{g_{tt}} \tilde{g_{\phi \phi}} \right) + \tilde{g_{tt}} x^{4},\\ 
e_{2} & = & - \frac{\left(a^{2} \tilde{g_{tt}} + a \tilde{g_{t\phi}} + \tilde{g_{tt}} x^{2}\right) \alpha}{e_{1} x \left(\tilde{g_{t\phi}}^2 - \tilde{g_{tt}} \tilde{g_{\phi \phi}}\right)}.
\end{eqnarray}

If we replace Eq. \eqref{efinal} into Eq. \eqref{lfinal}, and the latter into Eq. \eqref{derivada-pot}, after considerable algebra, we obtain an equation for the angular velocity $\tilde{\Omega}$:
\begin{equation}
B_{0} + B_{1} \sqrt{\beta} + B_{2} \beta= 0, 
\end{equation}
where,
\begin{eqnarray}
B_{0} & = & e_{1}^2 k_{1}^2 x^2 \left[ k_{2} \left(\tilde{g_{tt,x}} k_{2}-2
   \tilde{g_{t\phi,x}} k_{1}\right)+ \tilde{g_{\phi \phi,x}} k_{1}^2\right],\\
B_{1} & = & 2 k_{1}  e_{1} \alpha  k_{1}  \left(\tilde{a} \left(\tilde{a}
   k_{1}+k_{2}\right)- k_{1} x^2\right)\nonumber \\
  & + &  2 k_{1}  e_{1} k_{3} x^2 
   \left(\tilde{g_{t\phi,x}} k_{1} - \tilde{g_{tt,x}} k_{2}\right) \nonumber \\
   & + & 2 e_{1}^2 e_{2} k_{1} x^2 \left[k_{2} \left(\tilde{g_{tt,x}}
   k_{2}-2 \tilde{g_{t \phi,x}} k_{1}\right)+\tilde{g_{\phi \phi,x}} k_{1}^2\right],\\
B_{2} & = & e_{1}^2 e_{2}^2 x^2 \left[k_{2} \left(  \tilde{g_{tt,x}} k_{2}-2
   \tilde{g_{t\phi,x}} k_{1}\right)+ \tilde{g_{\phi \phi,x}} k_{1}^2\right] \nonumber \\
& + &  x^2 \left[k_{1}^2 \left(\tilde{g_{tt}} \tilde{g_{\phi \phi,x}}+\tilde{g_{tt,x}} \tilde{g_{\phi \phi
   }}\right)+\tilde{g_{tt,x}} k_{3}^2\right] \nonumber \\
 & - & 2 \tilde{g_{t\phi}} \tilde{g_{t\phi,x}}k_{1}^2 x^2 - 2 a k_{1} k_{3} \alpha \nonumber \\
 & + & 2 e_{1} e_{2} k_{3} x^2 \left(\tilde{g_{t\phi,x}} k_{1} - \tilde{g_{tt,x}} k_{2}\right) \nonumber \\
 & + & 2 e_{1} e_{2} k_{3} x^2  \alpha k_{1} \left[ a \left( a k_{1} + k_{2}  \right) - k_{1} x^{2}\right]. 
\end{eqnarray} 
and
\begin{eqnarray}
k_{2} & = &   \tilde{g_{t\phi}} + \tilde{\Omega} \ \tilde{g_{\phi \phi}} ,\\
k_{3} & = &  - \frac{  \alpha  \  \left[ -\left(x^2 + \tilde{a}^2\right) \ \tilde{\Omega}+\tilde{a}\right]}{x}.
\end{eqnarray}

Finally,
\begin{eqnarray}
\tilde{g_{tt}} & = & - \frac{\left(\tilde{\Delta} - {\tilde{a}}^{2}\right)}{x^2},\\
\tilde{g_{t\phi}} & = & \frac{\left[- \left(x^{2} +a^{2} \right) + \tilde{\Delta}\right] \tilde{a}}{x^{2}},\\
\tilde{g_{\phi \phi}} & = &   \frac{\left[ \left(x^{2} +a^{2} \right)^{2} -  \tilde{\Delta} \tilde{a}^{2}\right] }{x^{2}},\\
\tilde{g_{tt,x}} & = & - 2 \frac{\left(x-\alpha\right) \left(1 + \alpha\right)}{x^{3}},\\
\tilde{g_{t\phi,x}} & = &  2 \ a \frac{\left(x-\alpha\right) \left(1 + \alpha\right)}{x^{3}},\\
\tilde{g_{\phi \phi, x}} & = &  2 \frac{x^{4}- a^{2} \left(x-\alpha\right) \left(1+\alpha\right)}{x^{3}}.
\end{eqnarray}
Here, $\tilde{\Delta}$ is given by Eq. \eqref{delta}.

The heat $Q$ is related to the dimensionless function $f$ as \cite{pag+74}:
\begin{eqnarray}
Q(x) & = & \frac{c^3 \dot{M}}{4 \pi  \sqrt{- g}} f(x) = \frac{c^3 \dot{M}}{4 \pi c r_{g}^2 x^2}f(x)\\ 
& = & \frac{c^6 \dot{M}}{4 \pi {G_{N}}^2 M^2 x^2}f(x),\ \ \ \ \ \left[Q(x)\right] = {\rm erg \ cm^{-2} \ s^{-1}}.\nonumber
\end{eqnarray}
%The units of $Q(x)$ are erg cm$^{-2}$ s${-1}$.

If we assume that the disk is optically thick in the z-direction, every element of area on its surface radiates as a blackbody at the local effective surface temperature $T(x)$. By means of the Stefan Bolzmann's law, we obtain the temperature profile:
\begin{equation}
T(x) = z(x) \left(\frac{Q(x)}{\sigma}\right)^{1/4}, \ \ \ \ \left[T(x)\right] = {\rm K}.
\end{equation}
As usual, $\sigma$ denotes the Stefan-Boltzmann constant; the function $z(x)$ stands for the correction due to gravitational redshift and takes the form:
\begin{equation}
z(x) = \frac{\sqrt{x^{2}-2 (1+\alpha) x + a^2 + \alpha (1+\alpha)}}{x}. 
\end{equation}

Given the blackbody assumption, the emissivity per unit frecuency $I_{\nu}$ of each element of area on the disk is described by the Planck function:
\begin{equation}
I_{\nu}(\nu,x) = B_{\nu}(\nu,x) = \frac{2 h \nu^{3}}{c^{2} \left[e^{\left(h\nu/kT(x)\right)} -1\right]}.%\ \ \ \ \left[I_{\nu}(\nu,x)\right] = \frac{{\rm erg}} {{\rm s} \ {\rm cm}^{2} \ {\rm Hz}\ {\rm sr}}.
\end{equation}
Here $h$ and $k$ are the Planck and Boltzmann constants, respectively. The total luminosity at frecuency $\nu$  is:
\begin{eqnarray}
L_{\nu}(\nu) & = & \int_{x_{isco}}^{x_{out}} 2 \pi r_{g}^{2} I_{\nu} x \ dx, \\
& = & \frac{4 \pi h {G_{N}}^2 M^{2}  \  \nu^{3}}{c^{6}} \int_{x_{isco}}^{x_{out}} \frac{x \ dx}{\left[e^{\left(h\nu/kT(x)\right)} -1\right]}.%\ \ \ [L_{\nu}(\nu)] = {\rm erg}.\nonumber
\end{eqnarray}

We adopt for the radius of the outer edge of the disk \cite{dov+97} $r_{\mathrm{out}} = 11 r_{\mathrm{isco}}$.
 
We also investigate the structure of the accretion disk in the $z$-(vertical) direction in STVG theory. In the steady thin disk model developed by \cite{sha+73}, the typical scale-height of the disk in the $z$-direction is given by:
\begin{equation}\label{h}
H \cong  c_{\rm s} \left(\frac{r}{G_{\rm N} M}\right)^{1/2} r,
\end{equation}   
 where $c_{\rm s}$ is the sound speed. The latter expression was derived under the assumption of hydrostatic equilibrium in the $z$-direction.
 
For an accretion disk in STVG, expression \eqref{h} takes the form:
\begin{equation}
H_{\mathrm{STVG}} \cong  c_{\rm s} \left(\frac{r}{G_{\rm N} \left(1+\alpha \right)M}\right)^{1/2} r.
\end{equation} 

We estimate the deviation in scale-height $H$ of an accretion disk in STVG with respect to an accretion disk in Shakura $\&$ Sunyaev model as:
\begin{equation}
\frac{H_{\mathrm{STVG}}}{H}  \approx \left(1+\alpha\right)^{-1/2}.
\end{equation}
Hence, an accretion disk in STVG is thinner than in GR; this difference increases as the parameter $\alpha$ growths.

\subsection{Results}

\subsubsection{Primary cinematic parameters}\label{pcp}

\begin{figure}
\center
\includegraphics[angle=-90,width=8cm]{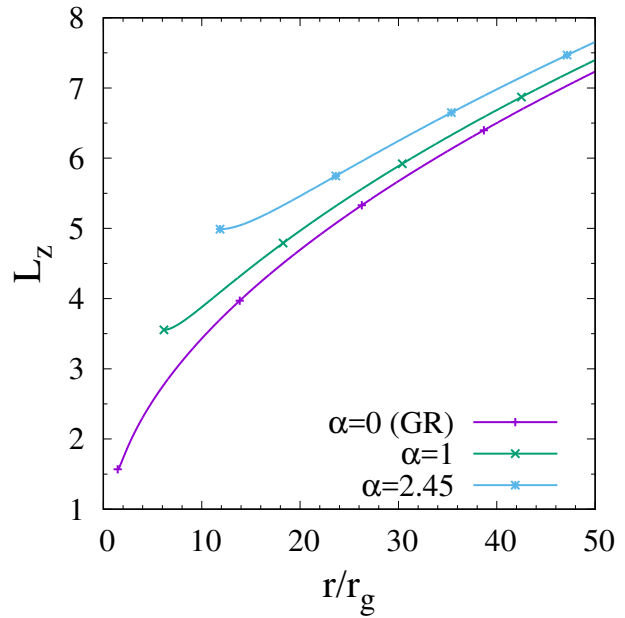}
%\resizebox{\hsize}{!}{\includegraphics{fig11.eps}}% Here is how to import EPS art
\caption{Plot of the specific angular momentum as a function of the radial coordinate for a test particle in a circular equatorial orbit around a supermassive Kerr STVG black hole of spin $\tilde{a} = 0.99$.}
 \label{am}
 \end{figure}

\begin{figure}
\center
\includegraphics[angle=-90,width=8cm]{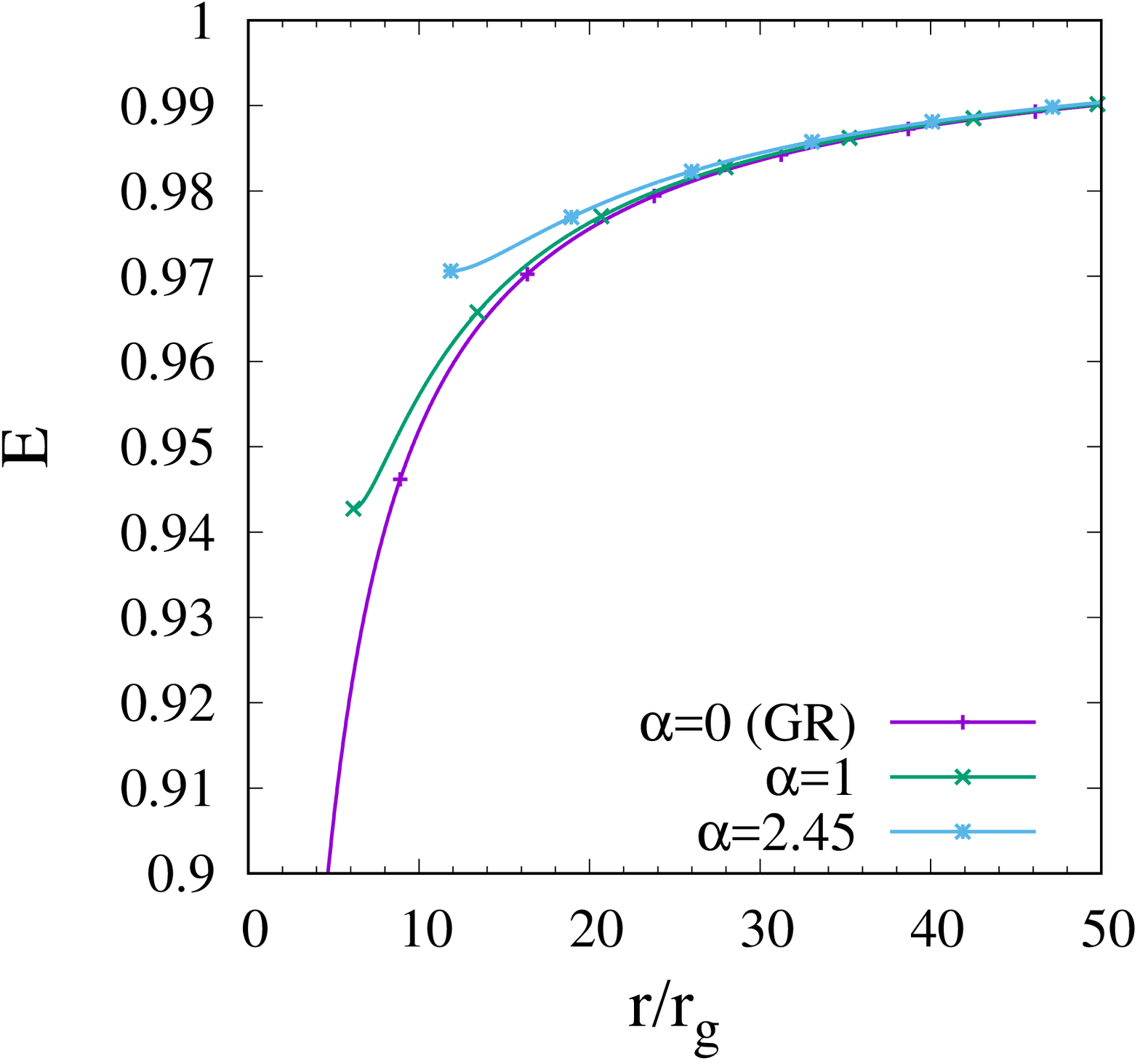}
%\resizebox{\hsize}{!}{\includegraphics{fig11.eps}}% Here is how to import EPS art
\caption{Plot of the specific energy as a function of the radial coordinate for a test particle in a circular equatorial orbit around a supermassive Kerr STVG black hole of spin $\tilde{a} = 0.99$.}
 \label{energy}
 \end{figure}

\begin{figure}
\center
\includegraphics[angle=-90,width=8cm]{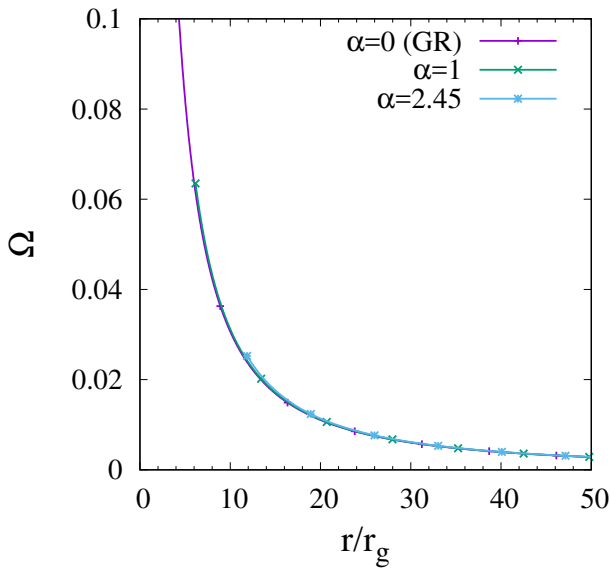}
%\resizebox{\hsize}{!}{\includegraphics{fig11.eps}}% Here is how to import EPS art
\caption{Plot of the angular velocity as a function of the radial coordinate for a test particle in a circular equatorial orbit around a super massive Kerr STVG black hole of spin $\tilde{a} = 0.99$.}
 \label{omega}
 \end{figure}

We have computed numerically the specific angular momentum $\tilde{L}$, the specific energy $\tilde{E}$, and angular velocity $\tilde{\Omega}$ for a test particle in a circular equatorial orbit around  Schwarzschild and Kerr STVG black holes, respectively; the spin of the Kerr-STVG black hole is $\tilde{a} = 0.99$. The most relevant results are display in Figs. \ref{am}, \ref{energy}, and \ref{omega} for the case of a supermassive Kerr black hole with $\alpha =1$, and $\alpha = 2.45$, respectively.  We also show for comparison the corresponding plots for a Kerr black hole of the same spin in GR.

We see that both $\tilde{L}$, and $\tilde{E}$ are larger in STVG than in GR. As the value of the parameter $\alpha$ increases the deviation with respect to GR is more prominent. On the contrary, there are no significant differences between the angular velocity in STVG and GR.

Notice that the slopes of the functions $\tilde{L}$ and $\tilde{E}$ in STVG (Figs. \ref{am} and \ref{energy}) are smaller than in GR. Such peculiarity in the behaviour of $\tilde{L}$ and $\tilde{E}$ becomes relevant in the calculation of the heat $Q$ produced by the accretion disk. From Eqs. \eqref{a0}, \eqref{a1}, and \eqref{a2}, we see that the coefficients of the differential equation \eqref{flux} for the function $f$ depend on the derivatives of $\tilde{L}$ and $\tilde{E}$. As we will show in the following subsections, the amount of heat radiated by an accretion disk in STVG is smaller than in GR. As a result, the temperature and luminosity distributions also decrease.

\subsubsection{Temperature and luminosity of a Schwarzschild-STVG black hole}

\begin{figure}
\center
\includegraphics[angle=-90,width=8cm]{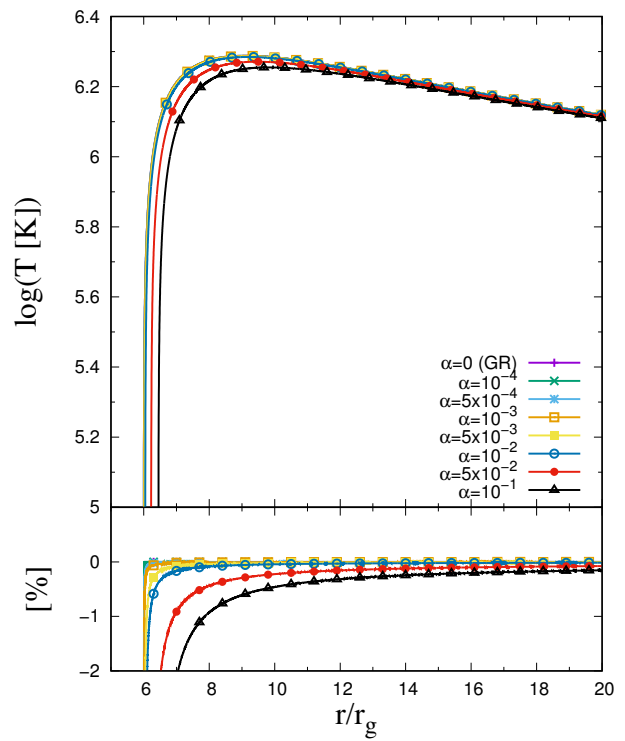}
%     \resizebox{\hsize}{!}{\includegraphics{fig5.eps}}% Here is how to import EPS art
\caption{\textit{Top}: Temperature as a function of the radial coordinate for a stellar Schwarzschild-STVG black hole. \textit{Bottom}: Residual plot of the temperature as a  function of the radial coordinate for a stellar Schwarzschild-STVG black hole.}
\label{fig5}
 \end{figure}

\begin{figure}
\center
\includegraphics[angle=-90,width=8cm]{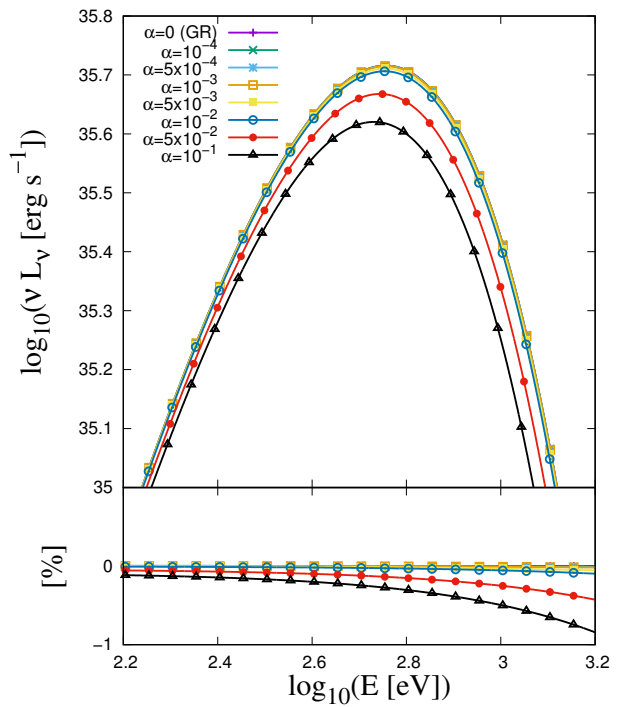}
%\resizebox{\hsize}{!}{\includegraphics{fig6.eps}}% Here is how to import EPS art
\caption{\textit{Top}: Luminosity as a a  function of the energy for a stellar Schwarzschild-STVG black hole. \textit{Bottom}: Residual plot of the luminosity as a  function of the energy for a stellar Schwarzschild-STVG black hole.}
 \label{fig6}
 \end{figure}

%\begin{figure*}
  %  \centering
    %\begin{subfigure}[b]{0.3\textwidth}
      %  \includegraphics[angle=-90,width=8cm]{fig-5.eps}
        %\caption{A gull}
        %\label{fig:gull}
   % \end{subfigure}
%    ~ %add desired spacing between images, e. g. ~, \quad, \qquad, \hfill etc. 
    %\begin{subfigure}[b]{0.3\textwidth}
    %    \includegraphics[angle=-90,width=8cm]{fig-6.eps}
        %\caption{A tiger}
        %\label{fig:tiger}
   % \end{subfigure}
 %   ~ %add desired spacing between images, e. g. ~, \quad, \qquad, \hfill etc. 
    %\begin{subfigure}[b]{0.3\textwidth}
       % \includegraphics[width=\textwidth]{mouse}
        %\caption{A mouse}
        %\label{fig:mouse}
    %\end{subfigure}
 %   \caption{Pictures of animals}\label{fig:animals}
%\end{figure*}

We show in Fig. \ref{fig5} the plot of the temperature as a function of the radial coordinate (above), and the corresponding temperature residual with GR (below). In Fig. \ref{fig6} we plot the luminosity as a function of the energy (above) and its corresponding residual (below) for a stellar-Schwarzschild-STVG black hole. We observe that the disk is colder as the parameter $\alpha$ increases. This feature is noticeable at the inner parts of the disk, while for large values of the radial coordinate, the value of the temperature tends to the corresponding value in GR. The deviations with repect to GR, however, are small: for $\alpha = 10^{-1}$ the largest difference in temperature is only of $0.2$ percent and $0.27$ percent in luminosity.

\begin{figure}
\center
\includegraphics[angle=-90,width=8cm]{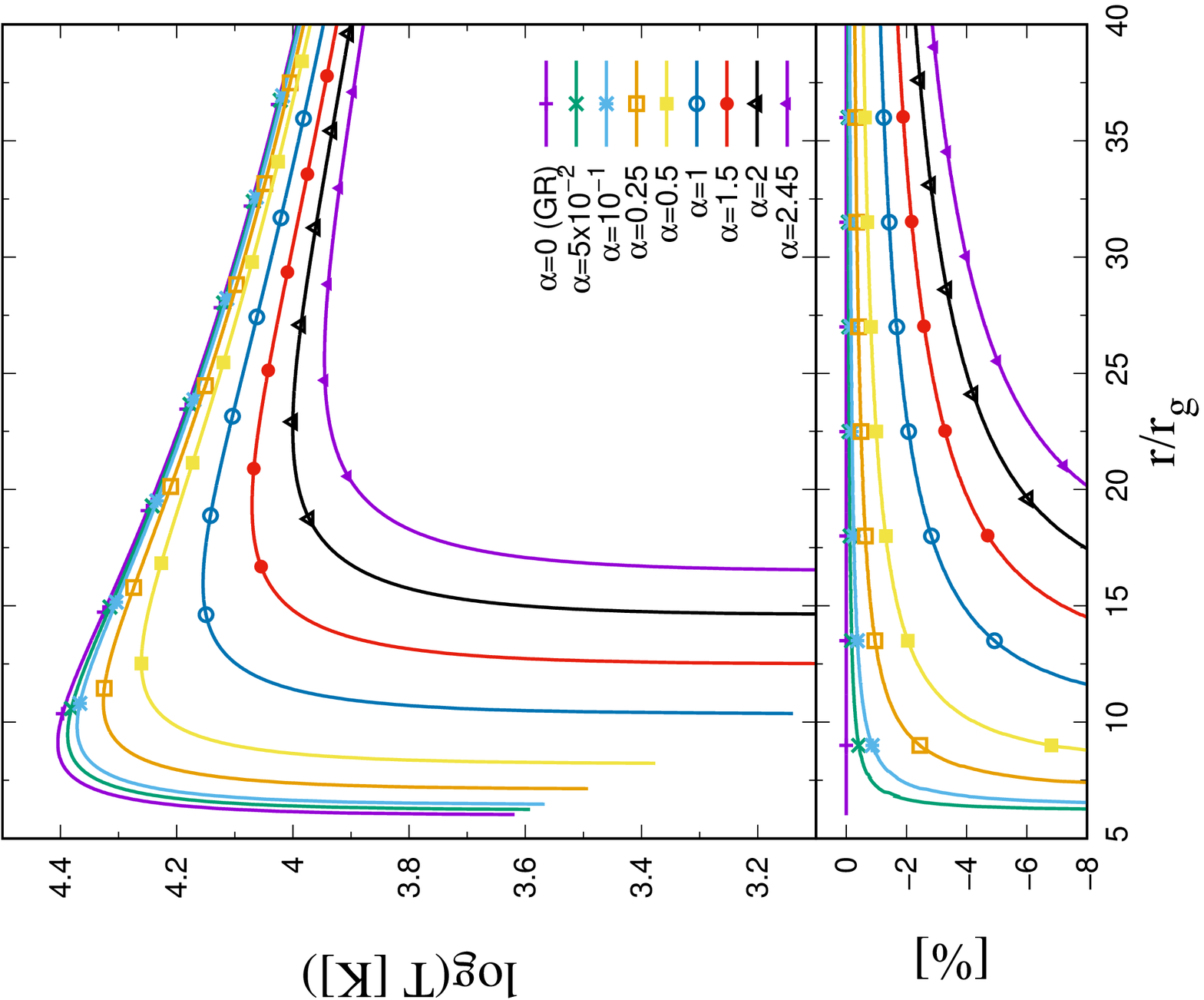}
%\resizebox{\hsize}{!}{\includegraphics{fig7.eps}}% Here is how to import EPS art
\caption{\textit{Top}: Temperature as a function of the radial coordinate for a supermassive Schwarzschild-STVG black hole. \textit{Bottom}: Residual plot of the temperature as a  function of the radial coordinate for a supermassive Schwarzschild-STVG black hole.}
 \label{fig7}
 \end{figure}

\begin{figure}
\center
\includegraphics[angle=-90,width=8cm]{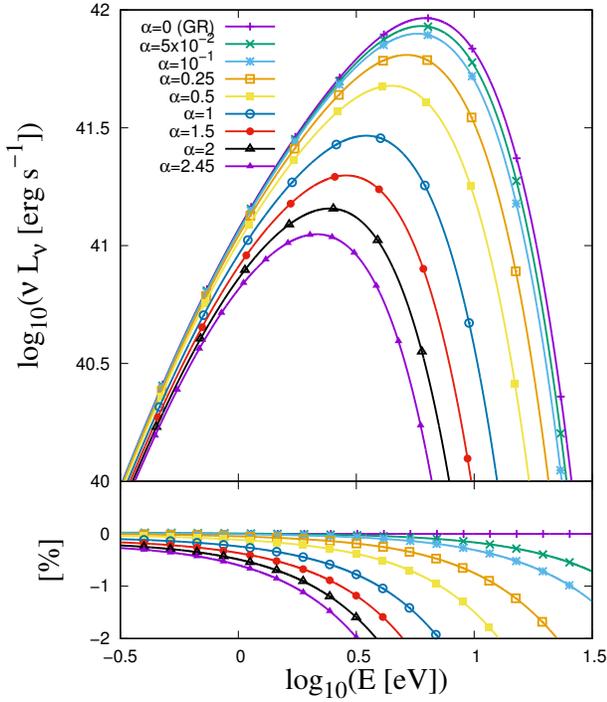}
%\resizebox{\hsize}{!}{\includegraphics{fig8.eps}}% Here is how to import EPS art
\caption{\textit{Top}: Luminosity as a  function of the energy for a supermassive Schwarzschild-STVG black hole. \textit{Bottom}: Residual plot of the luminosity as a  function of the energy for a supermassive Schwarzschild-STVG black hole.}
 \label{fig8}
 \end{figure}

For a supermassive black hole in Schwarzschild-STVG spacetime, the temperature of the inner parts of the disk is almost $3.5$ percent lower than in GR ($\alpha = 2.45$) (see Fig. \ref{fig7}). Accordingly, the peak of the emission decreases up to $2.18$ percent, and the corresponding energy is shifted toward lower energies (Fig. \ref{fig8}). 

%As shown in Section \ref{cine}, the radius of the ISCO for a stellar and supermassive Schwarzschild black hole in STVG is larger than the corresponding one in GR. 

\subsubsection{Temperature and luminosity for Kerr-STVG black hole}

\begin{figure}
\center
\includegraphics[angle=-90,width=8cm]{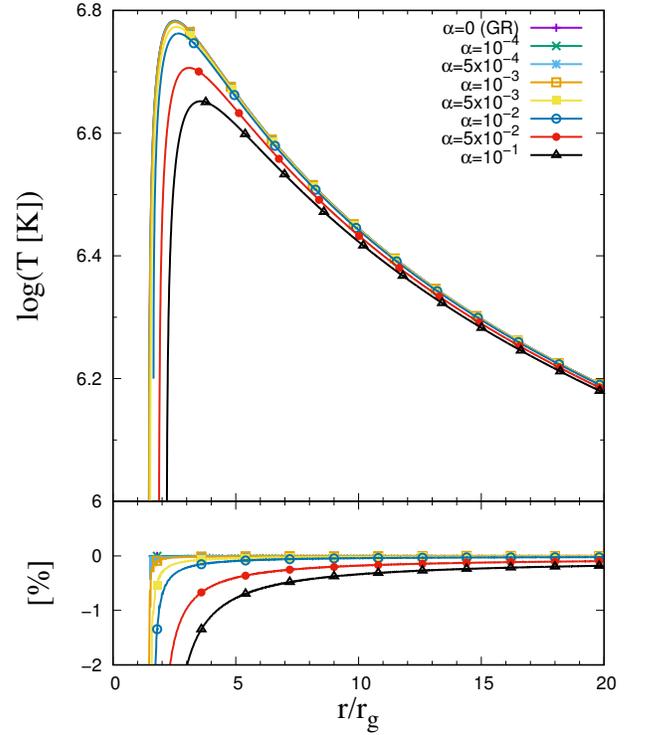}
%\resizebox{\hsize}{!}{\includegraphics{fig9.eps}}% Here is how to import EPS art
\caption{\textit{Top}: Temperature as a  function of the radial coordinate for a stellar Kerr-STVG black hole of spin $\tilde{a} = 0.99$. \textit{Bottom}: Residual plot of the temperature as a  function of the radial coordinate for a stellar Kerr-STVG black hole of spin $\tilde{a} = 0.99$.}
 \label{fig9}
 \end{figure}

\begin{figure}
\center
\includegraphics[angle=-90,width=8cm]{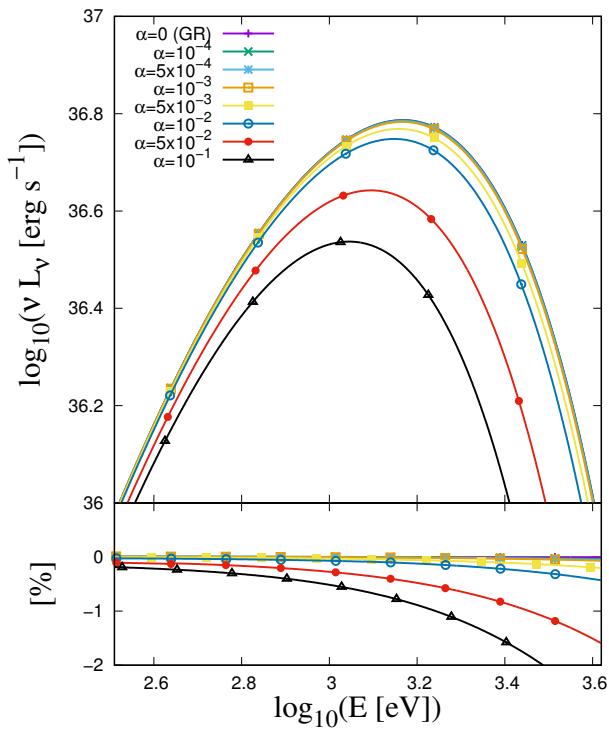}
%\resizebox{\hsize}{!}{\includegraphics{fig10.eps}}% Here is how to import EPS art
\caption{\textit{Top}: Luminosity as a  function of the energy for a stellar Kerr-STVG black hole of spin $\tilde{a} = 0.99$. \textit{Bottom}: Residual plot of the luminosity as a  function of the energy for a stellar Kerr-STVG black hole of spin $\tilde{a} = 0.99$.}
 \label{fig10}
 \end{figure}

We display in Figs. \ref{fig9} and \ref{fig10} the plot of the temperature as a function of the radial coordinate, and the plot of the luminosity as a function of the energy for a stellar Kerr-STVG black hole, respectively. The corresponding residual plots are shown at the botton of each figure. The accretion disk has a lower temperature than in GR; the largest difference in temperature is of $0.8$ percent for $\alpha = 10^{-1}$. The maximum luminosity of the disk also decreases in $0.67$ percent with respect to GR and the corresponding energy is shifted toward lower energies.

\begin{figure}
\center
\includegraphics[angle=-90,width=8cm]{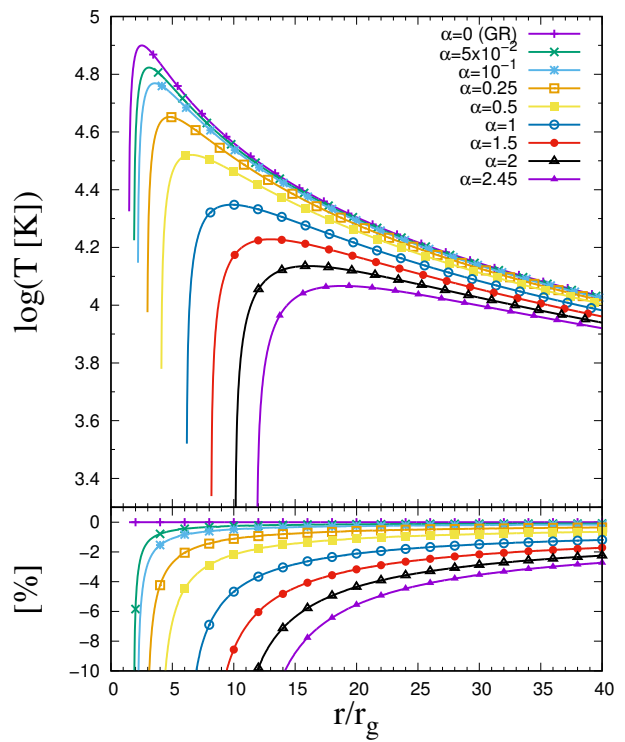}
%\resizebox{\hsize}{!}{\includegraphics{fig11.eps}}% Here is how to import EPS art
\caption{\textit{Top}: Temperature as a  function of the radial coordinate for a supermassive Kerr-STVG black hole of spin $\tilde{a} = 0.99$. \textit{Bottom}: Residual plot of the temperature as a function of the radial coordinate for a supermassive Kerr-STVG black hole of spin $\tilde{a} = 0.99$.}
 \label{fig11}
 \end{figure}

\begin{figure}
\center
\includegraphics[angle=-90,width=8cm]{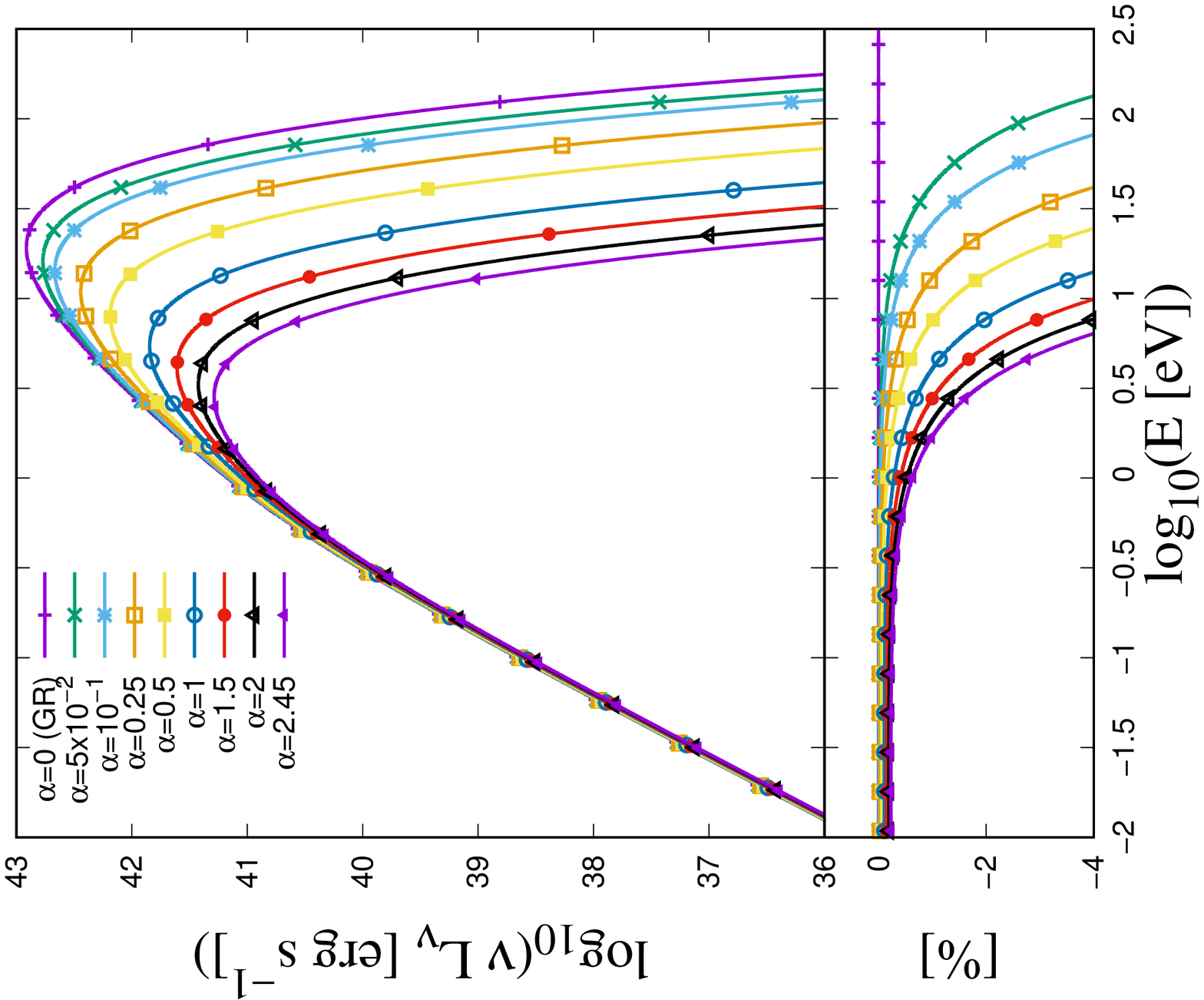}
%\resizebox{\hsize}{!}{\includegraphics{fig12.eps}}% Here is how to import EPS art
\caption{\textit{Top}: Luminosity as a  function of the energy for a supermassive Kerr-STVG black hole of spin $\tilde{a} = 0.99$. \textit{Bottom}: Residual plot of the luminosity as a  function of the energy for a supermassive Kerr-STVG black hole of spin $\tilde{a} = 0.99$.}
 \label{fig12}
 \end{figure}

The most significant deviations from GR arise for supermassive Kerr-STVG black holes, as shown in Figs. \ref{fig11} and \ref{fig12}. For $\alpha = 2.45$, the maximum temperature reduces up to $12.30$ percent, and the peak of the emission is almost $4$ percent lower than in GR.

\section{Discussion}\label{sec5}

Accretion disks in STVG around stellar and supermassive black holes are colder and underluminous than in GR. We can attribute two main reasons to these phenomena. First, the last stable circular orbit in Schwarzschild and Kerr-STVG black holes is farther from the hole than in GR; the higher the value of the parameter $\alpha$, the larger the radius of the ISCO. Since accretion disks get more luminous the closer they are to the hole, the contribution to the luminosity at high energies reduces for larger values of the parameter $\alpha$. Second, as already mention in Section \ref{pcp}, the calculation of the heat $Q$ of the disk depends on the derivatives of $\tilde{L}$ and $\tilde{E}$ which are smaller in STVG than in GR. This behavior is opposite to what is found in other modified theories of relativistic gravitation, such as $f(R)$, where disks tend to be sistematically hotter and more luminous than in GR \cite{per+13}.

We can understand why the ISCO in STVG is farther from the black hole than GR as follows: a) In the strong gravity regime, the gravitational repulsion due to the Yukawa-type force cannot counteract the enhanced gravitational attraction caused by the modified gravitational constant $G = G_{\rm N} (1+\alpha)$. As a result, the gravitational field is stronger than in GR, and the region of instability around the black hole increases displacing the position of the ISCO at larger radii. b) Because of the presence of the gravitational vector field $\phi^{\mu}$, and for rotating sources such as Kerr black holes, the vector potential in Boyer-Lindquist coordinates is:
\begin{equation}\label{a}
 \boldsymbol{\phi} = \frac{- Q r}{\rho^{2}} \left( \mathbf{dt} - a \sin^{2}{\theta} \ \mathbf{d\phi} \right).  
 \end{equation}
 The components of the four-velocity of a test particle in a circular equatorial orbit in STVG are:
 \begin{equation}\label{u}
 u^{\alpha} = \left(u^{t},0,0,u^{\phi}\right).  
 \end{equation}
Inserting Eqs. \eqref{a} and \eqref{u} into \eqref{eq-motion}, where $B^{\mu}_{\nu} = \partial^{\mu}A_{\nu} - \partial^{\nu}A_{\mu}$, a simple calculation shows that such a particle is subjected to gravitomagnetic Lorenz force in the radial direction, thus increasing the strenght of the gravitational field. Notice (see Section \ref{cine}) that the most significant deviations in the location of the ISCO with respect to GR are for supermassive Kerr black holes with $\alpha = 2.45$. The bigger the black hole, the stronger the deviation from GR. This provides a tool to observationally differenciate between both theories if reliable estimates of the black hole masses and luminosities are available

The spectral energy distributions obtained do not seem to contradict astronomical observations (see for example \cite{gou+11}). Thus, in the present work, no further limits can be imposed to the range of values of the parameter $\alpha$ that corresponds to stellar and supermassive black holes.

The observation of low luminosity Galactic nuclei, such as Sg $A^{\star}$, led to propose another accretion regime model different from the geometrically thin, optically thick accretion disk model, called Advection-Dominated Accretion Flows (ADAF). As STVG theory predicts thin relativistic accretion disks that are less luminous than in GR, it could be possible that a modified theory of gravitation could lead to the same observational predictions as ADAF in GR, making individual cases difficult to compare in the absence of independent information about the accretion regime.

\section{Conclusions}

We have studied stable circular orbits and relativistic accretion disks around Schwarzschild and Kerr black holes in the strong regime in STVG theory, for both stellar ($0 < \alpha < 0.1$) and supermassive sources ($0.03 < \alpha < 2.47$). We have found that the last stable circular orbit is larger than in GR. As the value of the parameter $\alpha$ increases, the ISCO is located farther from the black hole. This is due to the enhanced gravitational constant $G = G_{\rm N} (1+\alpha)$, and the presence of a gravitational radial Lorentz force that manifests in the strong regime.

In order to calculate the heat radiated by the accretion disk, we generalised the relativistic formula given by \cite{pag+74}, for the case in which external vector fields are present. We, then, computed the temperature and luminosity distributions for Schwarzschild and Kerr-STVG black holes, respectively. The disks are colder and underluminuous in comparison with thin relativistic accretion disks in GR. Accretion disks around supermassive Kerr STVG black holes present the largest deviations in temperature and luminosity. The spectral energy distributions predicted by STVG theory are not in contradiction with current astronomical observations given our ignorance of the details of the accretion regime.

Recently, Romero and collaborators \cite{rom+16} provided a criterion for the identification of some supermassive black hole binaries in which the less massive black hole carves an annular gap in the circumbinary disk. If the most massive of the two black holes lauches a relativistic jet, the spectral energy distribution (SED) at gamma ray energies has a unique signature that can be used to identify such systems. The distance between the black holes and hence, the location of the gap in the disk, will be different in STVG theory. This in turn implies that the SEDs at very high energies will differ from those obtained by \cite{rom+16}. Future observations of supermassive binary black hole systems with Cherenkov telescopes, such as the Cherenkov Telescope Array (CTA; \cite{act+11}), could be used to further constrain STVG theory in the strong gravity regime.

In the present paper, we have focused our investigations on the dynamical aspects of STVG as well as on the structure and emission in the continuous of thin relativistic accretion disks. More stringent constraints in the parameters of STVG can be achieved by studying quasi-periodic oscillations (QPOs), and the profile and equivalent width of the X-ray iron emission lines in black hole systems. We will adress such issues in a future work.

\begin{acknowledgments}
This work was supported by grants AYA2016-76012-C3-1-P (Ministerio de Educaci\'on, Cultura y Deporte, Espa\~{n}a) and PIP 0338 (CONICET, Argentina).     
  \end{acknowledgments}

\appendix

\section{Modified acceleration law in the weak field regime}

The equation of motion for material test particles, assuming $2GM/r \ll 1$ and the slow motion approximation $dr/ds \approx dr/dt \ll 1$ is \cite{mof06}:
\begin{equation}\label{eqmotion}
\frac{d^{2}r}{dt^{2}} - \frac{J_{\rm N}^{2}}{r^{3}} + \frac{GM}{r^{2}} = \frac{q}{m} \frac{d\phi_{0}}{dr},
\end{equation}
where $J_{\rm N}$ is the Newtonian orbital angular momentum, and $\phi_{0}$ is the $t$-component of the gravitational vector field $\phi_{\mu}$.

For weak gravitational fields to first order, the static equations for $\phi_{0}$ for the source-free case are (see Eq. \eqref{tb}):
\begin{equation}\label{47}
\vec{\nabla}^{2}\phi_{0} - \tilde{\mu}^{2} \phi_{0} = 0,
\end{equation}
where $\vec{\nabla}^{2}\phi_{0}$ is the Laplacian operator, and the contribution from the self-interaction potential $W(\phi)$ has been neglected. For a spherically symmetric static field $\phi_{0}$, Eq. \ref{47} takes the form:
\begin{equation}
\phi''_{0} + \frac{2}{r} \phi'_{0} - \tilde{\mu}^{2} \phi_{0} = 0,
\end{equation}
which has the Yukawa solution:
\begin{equation}
\phi_{0}(r) = - G_{\rm N} M \alpha \frac{\exp(-\tilde{\mu} \ r)}{\tilde{\mu} \ r},
\end{equation}
 If we replace the latter equation in Eq. \eqref{eqmotion}, we obtain:
\begin{equation}
\frac{d^{2}r}{dt^{2}} - \frac{J_{\rm N}^{2}}{r^{3}} + \frac{GM}{r^{2}} = G_{\rm N} M \alpha \frac{\exp(-\tilde{\mu} r)}{\tilde{\mu} r^{2}} \left(1 + \tilde{\mu} \ r \right).
\end{equation}
Then, the radial acceleration takes the form:
\begin{equation}\label{ace}
a(r) = - \frac{G_{\rm N}\left(1 + \alpha \right) \ M}{r^{2}} + G_{\rm N} M \alpha \frac{\exp(- \tilde{\mu} \ r) }{r^{2}} \left(1 +\tilde{\mu}  r\right),
\end{equation}
The first term of Eq. \eqref{ace} represents an enhanced gravitational attraction, and serves to explain galaxy rotation curves, light bending phenomena, and cosmological data, without dark matter. The second term represents gravitational repulsion and becomes relevant  when $ \tilde{\mu} \ r \ll 1$. This Yukawa-type force counteracts the enhanced attraction, and from the interplay, the Newtonian acceleration law is recovered at $ \tilde{\mu} \ r \ll 1$ scales.

%\nocite{*}

\bibliography{prdref}% Produces the bibliography via BibTeX.

\providecommand{\noopsort}[1]{}\providecommand{\singleletter}[1]{#1}%
\begin{thebibliography}{37}%
\makeatletter
\providecommand \@ifxundefined [1]{%
 \@ifx{#1\undefined}
}%
\providecommand \@ifnum [1]{%
 \ifnum #1\expandafter \@firstoftwo
 \else \expandafter \@secondoftwo
 \fi
}%
\providecommand \@ifx [1]{%
 \ifx #1\expandafter \@firstoftwo
 \else \expandafter \@secondoftwo
 \fi
}%
\providecommand \natexlab [1]{#1}%
\providecommand \enquote  [1]{``#1''}%
\providecommand \bibnamefont  [1]{#1}%
\providecommand \bibfnamefont [1]{#1}%
\providecommand \citenamefont [1]{#1}%
\providecommand \href@noop [0]{\@secondoftwo}%
\providecommand \href [0]{\begingroup \@sanitize@url \@href}%
\providecommand \@href[1]{\@@startlink{#1}\@@href}%
\providecommand \@@href[1]{\endgroup#1\@@endlink}%
\providecommand \@sanitize@url [0]{\catcode `\\12\catcode `\$12\catcode
  `\&12\catcode `\#12\catcode `\^12\catcode `\_12\catcode `\%12\relax}%
\providecommand \@@startlink[1]{}%
\providecommand \@@endlink[0]{}%
\providecommand \url  [0]{\begingroup\@sanitize@url \@url }%
\providecommand \@url [1]{\endgroup\@href {#1}{\urlprefix }}%
\providecommand \urlprefix  [0]{URL }%
\providecommand \Eprint [0]{\href }%
\providecommand \doibase [0]{http://dx.doi.org/}%
\providecommand \selectlanguage [0]{\@gobble}%
\providecommand \bibinfo  [0]{\@secondoftwo}%
\providecommand \bibfield  [0]{\@secondoftwo}%
\providecommand \translation [1]{[#1]}%
\providecommand \BibitemOpen [0]{}%
\providecommand \bibitemStop [0]{}%
\providecommand \bibitemNoStop [0]{.\EOS\space}%
\providecommand \EOS [0]{\spacefactor3000\relax}%
\providecommand \BibitemShut  [1]{\csname bibitem#1\endcsname}%
\let\auto@bib@innerbib\@empty
%</preamble>
\bibitem [{\citenamefont {{Einstein}}(1915)}]{ein15}%
  \BibitemOpen
  \bibfield  {author} {\bibinfo {author} {\bibfnamefont {A.}~\bibnamefont
  {{Einstein}}},\ }\href@noop {} {\bibfield  {journal} {\bibinfo  {journal}
  {Sitzungsberichte der K{\"o}niglich Preu{\ss}ischen Akademie der
  Wissenschaften (Berlin), Seite 844-847.}\ } (\bibinfo {year}
  {1915})}\BibitemShut {NoStop}%
\bibitem [{\citenamefont {{Abbott}}\ \emph {et~al.}(2016)\citenamefont
  {{Abbott}}, \citenamefont {{Abbott}}, \citenamefont {{Abbott}}, \citenamefont
  {{Abernathy}}, \citenamefont {{Acernese}}, \citenamefont {{Ackley}},
  \citenamefont {{Adams}}, \citenamefont {{Adams}}, \citenamefont {{Addesso}},
  \citenamefont {{Adhikari}},\ and\ \citenamefont {et~al.}}]{abb+16}%
  \BibitemOpen
  \bibfield  {author} {\bibinfo {author} {\bibfnamefont {B.~P.}\ \bibnamefont
  {{Abbott}}}, \bibinfo {author} {\bibfnamefont {R.}~\bibnamefont {{Abbott}}},
  \bibinfo {author} {\bibfnamefont {T.~D.}\ \bibnamefont {{Abbott}}}, \bibinfo
  {author} {\bibfnamefont {M.~R.}\ \bibnamefont {{Abernathy}}}, \bibinfo
  {author} {\bibfnamefont {F.}~\bibnamefont {{Acernese}}}, \bibinfo {author}
  {\bibfnamefont {K.}~\bibnamefont {{Ackley}}}, \bibinfo {author}
  {\bibfnamefont {C.}~\bibnamefont {{Adams}}}, \bibinfo {author} {\bibfnamefont
  {T.}~\bibnamefont {{Adams}}}, \bibinfo {author} {\bibfnamefont
  {P.}~\bibnamefont {{Addesso}}}, \bibinfo {author} {\bibfnamefont {R.~X.}\
  \bibnamefont {{Adhikari}}}, \ and\ \bibinfo {author} {\bibnamefont
  {et~al.}},\ }\href {\doibase 10.1103/PhysRevLett.116.061102} {\bibfield
  {journal} {\bibinfo  {journal} {Physical Review Letters}\ }\textbf {\bibinfo
  {volume} {116}},\ \bibinfo {eid} {061102} (\bibinfo {year} {2016})},\ \Eprint
  {http://arxiv.org/abs/1602.03837} {arXiv:1602.03837 [gr-qc]} \BibitemShut
  {NoStop}%
\bibitem [{\citenamefont {{Romero}}(2013)}]{rom13}%
  \BibitemOpen
  \bibfield  {author} {\bibinfo {author} {\bibfnamefont {G.~E.}\ \bibnamefont
  {{Romero}}},\ }\href@noop {} {\bibfield  {journal} {\bibinfo  {journal}
  {Foundations of Science}\ }\textbf {\bibinfo {volume} {18}},\ \bibinfo
  {pages} {297} (\bibinfo {year} {2013})},\ \Eprint
  {http://arxiv.org/abs/1210.2427} {arXiv:1210.2427 [physics.gen-ph]}
  \BibitemShut {NoStop}%
\bibitem [{\citenamefont {{Aprile}}\ \emph {et~al.}(2012)\citenamefont
  {{Aprile}}, \citenamefont {{Alfonsi}}, \citenamefont {{Arisaka}},
  \citenamefont {{Arneodo}}, \citenamefont {{Balan}}, \citenamefont {{Baudis}},
  \citenamefont {{Bauermeister}}, \citenamefont {{Behrens}}, \citenamefont
  {{Beltrame}}, \citenamefont {{Bokeloh}}, \citenamefont {{Brown}},
  \citenamefont {{Bruno}}, \citenamefont {{Budnik}}, \citenamefont {{Cardoso}},
  \citenamefont {{Chen}}, \citenamefont {{Choi}}, \citenamefont {{Cline}},
  \citenamefont {{Colijn}}, \citenamefont {{Contreras}}, \citenamefont
  {{Cussonneau}}, \citenamefont {{Decowski}}, \citenamefont {{Duchovni}},
  \citenamefont {{Fattori}}, \citenamefont {{Ferella}}, \citenamefont
  {{Fulgione}}, \citenamefont {{Gao}}, \citenamefont {{Garbini}}, \citenamefont
  {{Ghag}}, \citenamefont {{Giboni}}, \citenamefont {{Goetzke}}, \citenamefont
  {{Grignon}}, \citenamefont {{Gross}}, \citenamefont {{Hampel}}, \citenamefont
  {{Kaether}}, \citenamefont {{Kish}}, \citenamefont {{Lamblin}}, \citenamefont
  {{Landsman}}, \citenamefont {{Lang}}, \citenamefont {{Le Calloch}},
  \citenamefont {{Levy}}, \citenamefont {{Lim}}, \citenamefont {{Lin}},
  \citenamefont {{Lindemann}}, \citenamefont {{Lindner}}, \citenamefont
  {{Lopes}}, \citenamefont {{Lung}}, \citenamefont {{Marrod{\'a}n Undagoitia}},
  \citenamefont {{Massoli}}, \citenamefont {{Melgarejo Fernandez}},
  \citenamefont {{Meng}}, \citenamefont {{Molinario}}, \citenamefont {{Nativ}},
  \citenamefont {{Ni}}, \citenamefont {{Oberlack}}, \citenamefont {{Orrigo}},
  \citenamefont {{Pantic}}, \citenamefont {{Persiani}}, \citenamefont
  {{Plante}}, \citenamefont {{Priel}}, \citenamefont {{Rizzo}}, \citenamefont
  {{Rosendahl}}, \citenamefont {{dos Santos}}, \citenamefont {{Sartorelli}},
  \citenamefont {{Schreiner}}, \citenamefont {{Schumann}}, \citenamefont
  {{Scotto Lavina}}, \citenamefont {{Scovell}}, \citenamefont {{Selvi}},
  \citenamefont {{Shagin}}, \citenamefont {{Simgen}}, \citenamefont
  {{Teymourian}}, \citenamefont {{Thers}}, \citenamefont {{Vitells}},
  \citenamefont {{Wang}}, \citenamefont {{Weber}},\ and\ \citenamefont
  {{Weinheimer}}}]{apr+12}%
  \BibitemOpen
  \bibfield  {author} {\bibinfo {author} {\bibfnamefont {E.}~\bibnamefont
  {{Aprile}}}, \bibinfo {author} {\bibfnamefont {M.}~\bibnamefont {{Alfonsi}}},
  \bibinfo {author} {\bibfnamefont {K.}~\bibnamefont {{Arisaka}}}, \bibinfo
  {author} {\bibfnamefont {F.}~\bibnamefont {{Arneodo}}}, \bibinfo {author}
  {\bibfnamefont {C.}~\bibnamefont {{Balan}}}, \bibinfo {author} {\bibfnamefont
  {L.}~\bibnamefont {{Baudis}}}, \bibinfo {author} {\bibfnamefont
  {B.}~\bibnamefont {{Bauermeister}}}, \bibinfo {author} {\bibfnamefont
  {A.}~\bibnamefont {{Behrens}}}, \bibinfo {author} {\bibfnamefont
  {P.}~\bibnamefont {{Beltrame}}}, \bibinfo {author} {\bibfnamefont
  {K.}~\bibnamefont {{Bokeloh}}}, \bibinfo {author} {\bibfnamefont
  {E.}~\bibnamefont {{Brown}}}, \bibinfo {author} {\bibfnamefont
  {G.}~\bibnamefont {{Bruno}}}, \bibinfo {author} {\bibfnamefont
  {R.}~\bibnamefont {{Budnik}}}, \bibinfo {author} {\bibfnamefont {J.~M.~R.}\
  \bibnamefont {{Cardoso}}}, \bibinfo {author} {\bibfnamefont {W.-T.}\
  \bibnamefont {{Chen}}}, \bibinfo {author} {\bibfnamefont {B.}~\bibnamefont
  {{Choi}}}, \bibinfo {author} {\bibfnamefont {D.}~\bibnamefont {{Cline}}},
  \bibinfo {author} {\bibfnamefont {A.~P.}\ \bibnamefont {{Colijn}}}, \bibinfo
  {author} {\bibfnamefont {H.}~\bibnamefont {{Contreras}}}, \bibinfo {author}
  {\bibfnamefont {J.~P.}\ \bibnamefont {{Cussonneau}}}, \bibinfo {author}
  {\bibfnamefont {M.~P.}\ \bibnamefont {{Decowski}}}, \bibinfo {author}
  {\bibfnamefont {E.}~\bibnamefont {{Duchovni}}}, \bibinfo {author}
  {\bibfnamefont {S.}~\bibnamefont {{Fattori}}}, \bibinfo {author}
  {\bibfnamefont {A.~D.}\ \bibnamefont {{Ferella}}}, \bibinfo {author}
  {\bibfnamefont {W.}~\bibnamefont {{Fulgione}}}, \bibinfo {author}
  {\bibfnamefont {F.}~\bibnamefont {{Gao}}}, \bibinfo {author} {\bibfnamefont
  {M.}~\bibnamefont {{Garbini}}}, \bibinfo {author} {\bibfnamefont
  {C.}~\bibnamefont {{Ghag}}}, \bibinfo {author} {\bibfnamefont {K.-L.}\
  \bibnamefont {{Giboni}}}, \bibinfo {author} {\bibfnamefont {L.~W.}\
  \bibnamefont {{Goetzke}}}, \bibinfo {author} {\bibfnamefont {C.}~\bibnamefont
  {{Grignon}}}, \bibinfo {author} {\bibfnamefont {E.}~\bibnamefont {{Gross}}},
  \bibinfo {author} {\bibfnamefont {W.}~\bibnamefont {{Hampel}}}, \bibinfo
  {author} {\bibfnamefont {F.}~\bibnamefont {{Kaether}}}, \bibinfo {author}
  {\bibfnamefont {A.}~\bibnamefont {{Kish}}}, \bibinfo {author} {\bibfnamefont
  {J.}~\bibnamefont {{Lamblin}}}, \bibinfo {author} {\bibfnamefont
  {H.}~\bibnamefont {{Landsman}}}, \bibinfo {author} {\bibfnamefont {R.~F.}\
  \bibnamefont {{Lang}}}, \bibinfo {author} {\bibfnamefont {M.}~\bibnamefont
  {{Le Calloch}}}, \bibinfo {author} {\bibfnamefont {C.}~\bibnamefont
  {{Levy}}}, \bibinfo {author} {\bibfnamefont {K.~E.}\ \bibnamefont {{Lim}}},
  \bibinfo {author} {\bibfnamefont {Q.}~\bibnamefont {{Lin}}}, \bibinfo
  {author} {\bibfnamefont {S.}~\bibnamefont {{Lindemann}}}, \bibinfo {author}
  {\bibfnamefont {M.}~\bibnamefont {{Lindner}}}, \bibinfo {author}
  {\bibfnamefont {J.~A.~M.}\ \bibnamefont {{Lopes}}}, \bibinfo {author}
  {\bibfnamefont {K.}~\bibnamefont {{Lung}}}, \bibinfo {author} {\bibfnamefont
  {T.}~\bibnamefont {{Marrod{\'a}n Undagoitia}}}, \bibinfo {author}
  {\bibfnamefont {F.~V.}\ \bibnamefont {{Massoli}}}, \bibinfo {author}
  {\bibfnamefont {A.~J.}\ \bibnamefont {{Melgarejo Fernandez}}}, \bibinfo
  {author} {\bibfnamefont {Y.}~\bibnamefont {{Meng}}}, \bibinfo {author}
  {\bibfnamefont {A.}~\bibnamefont {{Molinario}}}, \bibinfo {author}
  {\bibfnamefont {E.}~\bibnamefont {{Nativ}}}, \bibinfo {author} {\bibfnamefont
  {K.}~\bibnamefont {{Ni}}}, \bibinfo {author} {\bibfnamefont {U.}~\bibnamefont
  {{Oberlack}}}, \bibinfo {author} {\bibfnamefont {S.~E.~A.}\ \bibnamefont
  {{Orrigo}}}, \bibinfo {author} {\bibfnamefont {E.}~\bibnamefont {{Pantic}}},
  \bibinfo {author} {\bibfnamefont {R.}~\bibnamefont {{Persiani}}}, \bibinfo
  {author} {\bibfnamefont {G.}~\bibnamefont {{Plante}}}, \bibinfo {author}
  {\bibfnamefont {N.}~\bibnamefont {{Priel}}}, \bibinfo {author} {\bibfnamefont
  {A.}~\bibnamefont {{Rizzo}}}, \bibinfo {author} {\bibfnamefont
  {S.}~\bibnamefont {{Rosendahl}}}, \bibinfo {author} {\bibfnamefont
  {J.~M.~F.}\ \bibnamefont {{dos Santos}}}, \bibinfo {author} {\bibfnamefont
  {G.}~\bibnamefont {{Sartorelli}}}, \bibinfo {author} {\bibfnamefont
  {J.}~\bibnamefont {{Schreiner}}}, \bibinfo {author} {\bibfnamefont
  {M.}~\bibnamefont {{Schumann}}}, \bibinfo {author} {\bibfnamefont
  {L.}~\bibnamefont {{Scotto Lavina}}}, \bibinfo {author} {\bibfnamefont
  {P.~R.}\ \bibnamefont {{Scovell}}}, \bibinfo {author} {\bibfnamefont
  {M.}~\bibnamefont {{Selvi}}}, \bibinfo {author} {\bibfnamefont
  {P.}~\bibnamefont {{Shagin}}}, \bibinfo {author} {\bibfnamefont
  {H.}~\bibnamefont {{Simgen}}}, \bibinfo {author} {\bibfnamefont
  {A.}~\bibnamefont {{Teymourian}}}, \bibinfo {author} {\bibfnamefont
  {D.}~\bibnamefont {{Thers}}}, \bibinfo {author} {\bibfnamefont
  {O.}~\bibnamefont {{Vitells}}}, \bibinfo {author} {\bibfnamefont
  {H.}~\bibnamefont {{Wang}}}, \bibinfo {author} {\bibfnamefont
  {M.}~\bibnamefont {{Weber}}}, \ and\ \bibinfo {author} {\bibfnamefont
  {C.}~\bibnamefont {{Weinheimer}}},\ }\href {\doibase
  10.1103/PhysRevLett.109.181301} {\bibfield  {journal} {\bibinfo  {journal}
  {Physical Review Letters}\ }\textbf {\bibinfo {volume} {109}},\ \bibinfo
  {eid} {181301} (\bibinfo {year} {2012})},\ \Eprint
  {http://arxiv.org/abs/1207.5988} {arXiv:1207.5988 [astro-ph.CO]} \BibitemShut
  {NoStop}%
\bibitem [{\citenamefont {{LUX Collaboration}}\ \emph
  {et~al.}(2013)\citenamefont {{LUX Collaboration}}, \citenamefont {{Akerib}},
  \citenamefont {{Araujo}}, \citenamefont {{Bai}}, \citenamefont {{Bailey}},
  \citenamefont {{Balajthy}}, \citenamefont {{Bedikian}}, \citenamefont
  {{Bernard}}, \citenamefont {{Bernstein}}, \citenamefont {{Bolozdynya}},
  \citenamefont {{Bradley}}, \citenamefont {{Byram}}, \citenamefont {{Cahn}},
  \citenamefont {{Carmona-Benitez}}, \citenamefont {{Chan}}, \citenamefont
  {{Chapman}}, \citenamefont {{Chiller}}, \citenamefont {{Chiller}},
  \citenamefont {{Clark}}, \citenamefont {{Coffey}}, \citenamefont {{Currie}},
  \citenamefont {{Curioni}}, \citenamefont {{Dazeley}}, \citenamefont {{de
  Viveiros}}, \citenamefont {{Dobi}}, \citenamefont {{Dobson}}, \citenamefont
  {{Dragowsky}}, \citenamefont {{Druszkiewicz}}, \citenamefont {{Edwards}},
  \citenamefont {{Faham}}, \citenamefont {{Fiorucci}}, \citenamefont
  {{Flores}}, \citenamefont {{Gaitskell}}, \citenamefont {{Gehman}},
  \citenamefont {{Ghag}}, \citenamefont {{Gibson}}, \citenamefont
  {{Gilchriese}}, \citenamefont {{Hall}}, \citenamefont {{Hanhardt}},
  \citenamefont {{Hertel}}, \citenamefont {{Horn}}, \citenamefont {{Huang}},
  \citenamefont {{Ihm}}, \citenamefont {{Jacobsen}}, \citenamefont {{Kastens}},
  \citenamefont {{Kazkaz}}, \citenamefont {{Knoche}}, \citenamefont {{Kyre}},
  \citenamefont {{Lander}}, \citenamefont {{Larsen}}, \citenamefont {{Lee}},
  \citenamefont {{Leonard}}, \citenamefont {{Lesko}}, \citenamefont
  {{Lindote}}, \citenamefont {{Lopes}}, \citenamefont {{Lyashenko}},
  \citenamefont {{Malling}}, \citenamefont {{Mannino}}, \citenamefont
  {{McKinsey}}, \citenamefont {{Mei}}, \citenamefont {{Mock}}, \citenamefont
  {{Moongweluwan}}, \citenamefont {{Morad}}, \citenamefont {{Morii}},
  \citenamefont {{Murphy}}, \citenamefont {{Nehrkorn}}, \citenamefont
  {{Nelson}}, \citenamefont {{Neves}}, \citenamefont {{Nikkel}}, \citenamefont
  {{Ott}}, \citenamefont {{Pangilinan}}, \citenamefont {{Parker}},
  \citenamefont {{Pease}}, \citenamefont {{Pech}}, \citenamefont {{Phelps}},
  \citenamefont {{Reichhart}}, \citenamefont {{Shutt}}, \citenamefont
  {{Silva}}, \citenamefont {{Skulski}}, \citenamefont {{Sofka}}, \citenamefont
  {{Solovov}}, \citenamefont {{Sorensen}}, \citenamefont {{Stiegler}},
  \citenamefont {{O`Sullivan}}, \citenamefont {{Sumner}}, \citenamefont
  {{Svoboda}}, \citenamefont {{Sweany}}, \citenamefont {{Szydagis}},
  \citenamefont {{Taylor}}, \citenamefont {{Tennyson}}, \citenamefont
  {{Tiedt}}, \citenamefont {{Tripathi}}, \citenamefont {{Uvarov}},
  \citenamefont {{Verbus}}, \citenamefont {{Walsh}}, \citenamefont {{Webb}},
  \citenamefont {{White}}, \citenamefont {{White}}, \citenamefont
  {{Witherell}}, \citenamefont {{Wlasenko}}, \citenamefont {{Wolfs}},
  \citenamefont {{Woods}},\ and\ \citenamefont {{Zhang}}}]{lux+13}%
  \BibitemOpen
  \bibfield  {author} {\bibinfo {author} {\bibnamefont {{LUX Collaboration}}},
  \bibinfo {author} {\bibfnamefont {D.~S.}\ \bibnamefont {{Akerib}}}, \bibinfo
  {author} {\bibfnamefont {H.~M.}\ \bibnamefont {{Araujo}}}, \bibinfo {author}
  {\bibfnamefont {X.}~\bibnamefont {{Bai}}}, \bibinfo {author} {\bibfnamefont
  {A.~J.}\ \bibnamefont {{Bailey}}}, \bibinfo {author} {\bibfnamefont
  {J.}~\bibnamefont {{Balajthy}}}, \bibinfo {author} {\bibfnamefont
  {S.}~\bibnamefont {{Bedikian}}}, \bibinfo {author} {\bibfnamefont
  {E.}~\bibnamefont {{Bernard}}}, \bibinfo {author} {\bibfnamefont
  {A.}~\bibnamefont {{Bernstein}}}, \bibinfo {author} {\bibfnamefont
  {A.}~\bibnamefont {{Bolozdynya}}}, \bibinfo {author} {\bibfnamefont
  {A.}~\bibnamefont {{Bradley}}}, \bibinfo {author} {\bibfnamefont
  {D.}~\bibnamefont {{Byram}}}, \bibinfo {author} {\bibfnamefont {S.~B.}\
  \bibnamefont {{Cahn}}}, \bibinfo {author} {\bibfnamefont {M.~C.}\
  \bibnamefont {{Carmona-Benitez}}}, \bibinfo {author} {\bibfnamefont
  {C.}~\bibnamefont {{Chan}}}, \bibinfo {author} {\bibfnamefont {J.~J.}\
  \bibnamefont {{Chapman}}}, \bibinfo {author} {\bibfnamefont {A.~A.}\
  \bibnamefont {{Chiller}}}, \bibinfo {author} {\bibfnamefont {C.}~\bibnamefont
  {{Chiller}}}, \bibinfo {author} {\bibfnamefont {K.}~\bibnamefont {{Clark}}},
  \bibinfo {author} {\bibfnamefont {T.}~\bibnamefont {{Coffey}}}, \bibinfo
  {author} {\bibfnamefont {A.}~\bibnamefont {{Currie}}}, \bibinfo {author}
  {\bibfnamefont {A.}~\bibnamefont {{Curioni}}}, \bibinfo {author}
  {\bibfnamefont {S.}~\bibnamefont {{Dazeley}}}, \bibinfo {author}
  {\bibfnamefont {L.}~\bibnamefont {{de Viveiros}}}, \bibinfo {author}
  {\bibfnamefont {A.}~\bibnamefont {{Dobi}}}, \bibinfo {author} {\bibfnamefont
  {J.}~\bibnamefont {{Dobson}}}, \bibinfo {author} {\bibfnamefont {E.~M.}\
  \bibnamefont {{Dragowsky}}}, \bibinfo {author} {\bibfnamefont
  {E.}~\bibnamefont {{Druszkiewicz}}}, \bibinfo {author} {\bibfnamefont
  {B.}~\bibnamefont {{Edwards}}}, \bibinfo {author} {\bibfnamefont {C.~H.}\
  \bibnamefont {{Faham}}}, \bibinfo {author} {\bibfnamefont {S.}~\bibnamefont
  {{Fiorucci}}}, \bibinfo {author} {\bibfnamefont {C.}~\bibnamefont
  {{Flores}}}, \bibinfo {author} {\bibfnamefont {R.~J.}\ \bibnamefont
  {{Gaitskell}}}, \bibinfo {author} {\bibfnamefont {V.~M.}\ \bibnamefont
  {{Gehman}}}, \bibinfo {author} {\bibfnamefont {C.}~\bibnamefont {{Ghag}}},
  \bibinfo {author} {\bibfnamefont {K.~R.}\ \bibnamefont {{Gibson}}}, \bibinfo
  {author} {\bibfnamefont {M.~G.~D.}\ \bibnamefont {{Gilchriese}}}, \bibinfo
  {author} {\bibfnamefont {C.}~\bibnamefont {{Hall}}}, \bibinfo {author}
  {\bibfnamefont {M.}~\bibnamefont {{Hanhardt}}}, \bibinfo {author}
  {\bibfnamefont {S.~A.}\ \bibnamefont {{Hertel}}}, \bibinfo {author}
  {\bibfnamefont {M.}~\bibnamefont {{Horn}}}, \bibinfo {author} {\bibfnamefont
  {D.~Q.}\ \bibnamefont {{Huang}}}, \bibinfo {author} {\bibfnamefont
  {M.}~\bibnamefont {{Ihm}}}, \bibinfo {author} {\bibfnamefont {R.~G.}\
  \bibnamefont {{Jacobsen}}}, \bibinfo {author} {\bibfnamefont
  {L.}~\bibnamefont {{Kastens}}}, \bibinfo {author} {\bibfnamefont
  {K.}~\bibnamefont {{Kazkaz}}}, \bibinfo {author} {\bibfnamefont
  {R.}~\bibnamefont {{Knoche}}}, \bibinfo {author} {\bibfnamefont
  {S.}~\bibnamefont {{Kyre}}}, \bibinfo {author} {\bibfnamefont
  {R.}~\bibnamefont {{Lander}}}, \bibinfo {author} {\bibfnamefont {N.~A.}\
  \bibnamefont {{Larsen}}}, \bibinfo {author} {\bibfnamefont {C.}~\bibnamefont
  {{Lee}}}, \bibinfo {author} {\bibfnamefont {D.~S.}\ \bibnamefont
  {{Leonard}}}, \bibinfo {author} {\bibfnamefont {K.~T.}\ \bibnamefont
  {{Lesko}}}, \bibinfo {author} {\bibfnamefont {A.}~\bibnamefont {{Lindote}}},
  \bibinfo {author} {\bibfnamefont {M.~I.}\ \bibnamefont {{Lopes}}}, \bibinfo
  {author} {\bibfnamefont {A.}~\bibnamefont {{Lyashenko}}}, \bibinfo {author}
  {\bibfnamefont {D.~C.}\ \bibnamefont {{Malling}}}, \bibinfo {author}
  {\bibfnamefont {R.}~\bibnamefont {{Mannino}}}, \bibinfo {author}
  {\bibfnamefont {D.~N.}\ \bibnamefont {{McKinsey}}}, \bibinfo {author}
  {\bibfnamefont {D.-M.}\ \bibnamefont {{Mei}}}, \bibinfo {author}
  {\bibfnamefont {J.}~\bibnamefont {{Mock}}}, \bibinfo {author} {\bibfnamefont
  {M.}~\bibnamefont {{Moongweluwan}}}, \bibinfo {author} {\bibfnamefont
  {J.}~\bibnamefont {{Morad}}}, \bibinfo {author} {\bibfnamefont
  {M.}~\bibnamefont {{Morii}}}, \bibinfo {author} {\bibfnamefont {A.~S.~J.}\
  \bibnamefont {{Murphy}}}, \bibinfo {author} {\bibfnamefont {C.}~\bibnamefont
  {{Nehrkorn}}}, \bibinfo {author} {\bibfnamefont {H.}~\bibnamefont
  {{Nelson}}}, \bibinfo {author} {\bibfnamefont {F.}~\bibnamefont {{Neves}}},
  \bibinfo {author} {\bibfnamefont {J.~A.}\ \bibnamefont {{Nikkel}}}, \bibinfo
  {author} {\bibfnamefont {R.~A.}\ \bibnamefont {{Ott}}}, \bibinfo {author}
  {\bibfnamefont {M.}~\bibnamefont {{Pangilinan}}}, \bibinfo {author}
  {\bibfnamefont {P.~D.}\ \bibnamefont {{Parker}}}, \bibinfo {author}
  {\bibfnamefont {E.~K.}\ \bibnamefont {{Pease}}}, \bibinfo {author}
  {\bibfnamefont {K.}~\bibnamefont {{Pech}}}, \bibinfo {author} {\bibfnamefont
  {P.}~\bibnamefont {{Phelps}}}, \bibinfo {author} {\bibfnamefont
  {L.}~\bibnamefont {{Reichhart}}}, \bibinfo {author} {\bibfnamefont
  {T.}~\bibnamefont {{Shutt}}}, \bibinfo {author} {\bibfnamefont
  {C.}~\bibnamefont {{Silva}}}, \bibinfo {author} {\bibfnamefont
  {W.}~\bibnamefont {{Skulski}}}, \bibinfo {author} {\bibfnamefont {C.~J.}\
  \bibnamefont {{Sofka}}}, \bibinfo {author} {\bibfnamefont {V.~N.}\
  \bibnamefont {{Solovov}}}, \bibinfo {author} {\bibfnamefont {P.}~\bibnamefont
  {{Sorensen}}}, \bibinfo {author} {\bibfnamefont {T.}~\bibnamefont
  {{Stiegler}}}, \bibinfo {author} {\bibfnamefont {K.}~\bibnamefont
  {{O`Sullivan}}}, \bibinfo {author} {\bibfnamefont {T.~J.}\ \bibnamefont
  {{Sumner}}}, \bibinfo {author} {\bibfnamefont {R.}~\bibnamefont {{Svoboda}}},
  \bibinfo {author} {\bibfnamefont {M.}~\bibnamefont {{Sweany}}}, \bibinfo
  {author} {\bibfnamefont {M.}~\bibnamefont {{Szydagis}}}, \bibinfo {author}
  {\bibfnamefont {D.}~\bibnamefont {{Taylor}}}, \bibinfo {author}
  {\bibfnamefont {B.}~\bibnamefont {{Tennyson}}}, \bibinfo {author}
  {\bibfnamefont {D.~R.}\ \bibnamefont {{Tiedt}}}, \bibinfo {author}
  {\bibfnamefont {M.}~\bibnamefont {{Tripathi}}}, \bibinfo {author}
  {\bibfnamefont {S.}~\bibnamefont {{Uvarov}}}, \bibinfo {author}
  {\bibfnamefont {J.~R.}\ \bibnamefont {{Verbus}}}, \bibinfo {author}
  {\bibfnamefont {N.}~\bibnamefont {{Walsh}}}, \bibinfo {author} {\bibfnamefont
  {R.}~\bibnamefont {{Webb}}}, \bibinfo {author} {\bibfnamefont {J.~T.}\
  \bibnamefont {{White}}}, \bibinfo {author} {\bibfnamefont {D.}~\bibnamefont
  {{White}}}, \bibinfo {author} {\bibfnamefont {M.~S.}\ \bibnamefont
  {{Witherell}}}, \bibinfo {author} {\bibfnamefont {M.}~\bibnamefont
  {{Wlasenko}}}, \bibinfo {author} {\bibfnamefont {F.~L.~H.}\ \bibnamefont
  {{Wolfs}}}, \bibinfo {author} {\bibfnamefont {M.}~\bibnamefont {{Woods}}}, \
  and\ \bibinfo {author} {\bibfnamefont {C.}~\bibnamefont {{Zhang}}},\
  }\href@noop {} {\bibfield  {journal} {\bibinfo  {journal} {ArXiv e-prints}\ }
  (\bibinfo {year} {2013})},\ \Eprint {http://arxiv.org/abs/1310.8214}
  {arXiv:1310.8214 [astro-ph.CO]} \BibitemShut {NoStop}%
\bibitem [{\citenamefont {{Agnese}}\ \emph {et~al.}(2014)\citenamefont
  {{Agnese}}, \citenamefont {{Anderson}}, \citenamefont {{Asai}}, \citenamefont
  {{Balakishiyeva}}, \citenamefont {{Basu Thakur}}, \citenamefont {{Bauer}},
  \citenamefont {{Beaty}}, \citenamefont {{Billard}}, \citenamefont
  {{Borgland}}, \citenamefont {{Bowles}}, \citenamefont {{Brandt}},
  \citenamefont {{Brink}}, \citenamefont {{Bunker}}, \citenamefont {{Cabrera}},
  \citenamefont {{Caldwell}}, \citenamefont {{Cerdeno}}, \citenamefont
  {{Chagani}}, \citenamefont {{Chen}}, \citenamefont {{Cherry}}, \citenamefont
  {{Cooley}}, \citenamefont {{Cornell}}, \citenamefont {{Crewdson}},
  \citenamefont {{Cushman}}, \citenamefont {{Daal}}, \citenamefont {{DeVaney}},
  \citenamefont {{Di Stefano}}, \citenamefont {{Silva}}, \citenamefont
  {{Doughty}}, \citenamefont {{Esteban}}, \citenamefont {{Fallows}},
  \citenamefont {{Figueroa-Feliciano}}, \citenamefont {{Godfrey}},
  \citenamefont {{Golwala}}, \citenamefont {{Hall}}, \citenamefont {{Hansen}},
  \citenamefont {{Harris}}, \citenamefont {{Hertel}}, \citenamefont {{Hines}},
  \citenamefont {{Hofer}}, \citenamefont {{Holmgren}}, \citenamefont {{Hsu}},
  \citenamefont {{Huber}}, \citenamefont {{Jastram}}, \citenamefont {{Kamaev}},
  \citenamefont {{Kara}}, \citenamefont {{Kelsey}}, \citenamefont {{Kenany}},
  \citenamefont {{Kennedy}}, \citenamefont {{Kiveni}}, \citenamefont {{Koch}},
  \citenamefont {{Leder}}, \citenamefont {{Loer}}, \citenamefont {{Lopez
  Asamar}}, \citenamefont {{Mahapatra}}, \citenamefont {{Mandic}},
  \citenamefont {{Martinez}}, \citenamefont {{McCarthy}}, \citenamefont
  {{Mirabolfathi}}, \citenamefont {{Moffatt}}, \citenamefont {{Nelson}},
  \citenamefont {{Novak}}, \citenamefont {{Page}}, \citenamefont {{Partridge}},
  \citenamefont {{Pepin}}, \citenamefont {{Phipps}}, \citenamefont {{Platt}},
  \citenamefont {{Prasad}}, \citenamefont {{Pyle}}, \citenamefont {{Qiu}},
  \citenamefont {{Rau}}, \citenamefont {{Redl}}, \citenamefont {{Reisetter}},
  \citenamefont {{Resch}}, \citenamefont {{Ricci}}, \citenamefont {{Ruschman}},
  \citenamefont {{Saab}}, \citenamefont {{Sadoulet}}, \citenamefont {{Sander}},
  \citenamefont {{Schmitt}}, \citenamefont {{Schneck}}, \citenamefont
  {{Schnee}}, \citenamefont {{Scorza}}, \citenamefont {{Seitz}}, \citenamefont
  {{Serfass}}, \citenamefont {{Shank}}, \citenamefont {{Speller}},
  \citenamefont {{Tomada}}, \citenamefont {{Upadhyayula}}, \citenamefont
  {{Villano}}, \citenamefont {{Welliver}}, \citenamefont {{Wright}},
  \citenamefont {{Yellin}}, \citenamefont {{Yen}}, \citenamefont {{Young}},
  \citenamefont {{Zhang}},\ and\ \citenamefont {{SuperCDMS
  Collaboration}}}]{agn+14}%
  \BibitemOpen
  \bibfield  {author} {\bibinfo {author} {\bibfnamefont {R.}~\bibnamefont
  {{Agnese}}}, \bibinfo {author} {\bibfnamefont {A.~J.}\ \bibnamefont
  {{Anderson}}}, \bibinfo {author} {\bibfnamefont {M.}~\bibnamefont {{Asai}}},
  \bibinfo {author} {\bibfnamefont {D.}~\bibnamefont {{Balakishiyeva}}},
  \bibinfo {author} {\bibfnamefont {R.}~\bibnamefont {{Basu Thakur}}}, \bibinfo
  {author} {\bibfnamefont {D.~A.}\ \bibnamefont {{Bauer}}}, \bibinfo {author}
  {\bibfnamefont {J.}~\bibnamefont {{Beaty}}}, \bibinfo {author} {\bibfnamefont
  {J.}~\bibnamefont {{Billard}}}, \bibinfo {author} {\bibfnamefont
  {A.}~\bibnamefont {{Borgland}}}, \bibinfo {author} {\bibfnamefont {M.~A.}\
  \bibnamefont {{Bowles}}}, \bibinfo {author} {\bibfnamefont {D.}~\bibnamefont
  {{Brandt}}}, \bibinfo {author} {\bibfnamefont {P.~L.}\ \bibnamefont
  {{Brink}}}, \bibinfo {author} {\bibfnamefont {R.}~\bibnamefont {{Bunker}}},
  \bibinfo {author} {\bibfnamefont {B.}~\bibnamefont {{Cabrera}}}, \bibinfo
  {author} {\bibfnamefont {D.~O.}\ \bibnamefont {{Caldwell}}}, \bibinfo
  {author} {\bibfnamefont {D.~G.}\ \bibnamefont {{Cerdeno}}}, \bibinfo {author}
  {\bibfnamefont {H.}~\bibnamefont {{Chagani}}}, \bibinfo {author}
  {\bibfnamefont {Y.}~\bibnamefont {{Chen}}}, \bibinfo {author} {\bibfnamefont
  {M.}~\bibnamefont {{Cherry}}}, \bibinfo {author} {\bibfnamefont
  {J.}~\bibnamefont {{Cooley}}}, \bibinfo {author} {\bibfnamefont
  {B.}~\bibnamefont {{Cornell}}}, \bibinfo {author} {\bibfnamefont {C.~H.}\
  \bibnamefont {{Crewdson}}}, \bibinfo {author} {\bibfnamefont
  {P.}~\bibnamefont {{Cushman}}}, \bibinfo {author} {\bibfnamefont
  {M.}~\bibnamefont {{Daal}}}, \bibinfo {author} {\bibfnamefont
  {D.}~\bibnamefont {{DeVaney}}}, \bibinfo {author} {\bibfnamefont {P.~C.~F.}\
  \bibnamefont {{Di Stefano}}}, \bibinfo {author} {\bibfnamefont {E.~D.~C.~E.}\
  \bibnamefont {{Silva}}}, \bibinfo {author} {\bibfnamefont {T.}~\bibnamefont
  {{Doughty}}}, \bibinfo {author} {\bibfnamefont {L.}~\bibnamefont
  {{Esteban}}}, \bibinfo {author} {\bibfnamefont {S.}~\bibnamefont
  {{Fallows}}}, \bibinfo {author} {\bibfnamefont {E.}~\bibnamefont
  {{Figueroa-Feliciano}}}, \bibinfo {author} {\bibfnamefont {G.~L.}\
  \bibnamefont {{Godfrey}}}, \bibinfo {author} {\bibfnamefont {S.~R.}\
  \bibnamefont {{Golwala}}}, \bibinfo {author} {\bibfnamefont {J.}~\bibnamefont
  {{Hall}}}, \bibinfo {author} {\bibfnamefont {S.}~\bibnamefont {{Hansen}}},
  \bibinfo {author} {\bibfnamefont {H.~R.}\ \bibnamefont {{Harris}}}, \bibinfo
  {author} {\bibfnamefont {S.~A.}\ \bibnamefont {{Hertel}}}, \bibinfo {author}
  {\bibfnamefont {B.~A.}\ \bibnamefont {{Hines}}}, \bibinfo {author}
  {\bibfnamefont {T.}~\bibnamefont {{Hofer}}}, \bibinfo {author} {\bibfnamefont
  {D.}~\bibnamefont {{Holmgren}}}, \bibinfo {author} {\bibfnamefont
  {L.}~\bibnamefont {{Hsu}}}, \bibinfo {author} {\bibfnamefont {M.~E.}\
  \bibnamefont {{Huber}}}, \bibinfo {author} {\bibfnamefont {A.}~\bibnamefont
  {{Jastram}}}, \bibinfo {author} {\bibfnamefont {O.}~\bibnamefont {{Kamaev}}},
  \bibinfo {author} {\bibfnamefont {B.}~\bibnamefont {{Kara}}}, \bibinfo
  {author} {\bibfnamefont {M.~H.}\ \bibnamefont {{Kelsey}}}, \bibinfo {author}
  {\bibfnamefont {S.}~\bibnamefont {{Kenany}}}, \bibinfo {author}
  {\bibfnamefont {A.}~\bibnamefont {{Kennedy}}}, \bibinfo {author}
  {\bibfnamefont {M.}~\bibnamefont {{Kiveni}}}, \bibinfo {author}
  {\bibfnamefont {K.}~\bibnamefont {{Koch}}}, \bibinfo {author} {\bibfnamefont
  {A.}~\bibnamefont {{Leder}}}, \bibinfo {author} {\bibfnamefont
  {B.}~\bibnamefont {{Loer}}}, \bibinfo {author} {\bibfnamefont
  {E.}~\bibnamefont {{Lopez Asamar}}}, \bibinfo {author} {\bibfnamefont
  {R.}~\bibnamefont {{Mahapatra}}}, \bibinfo {author} {\bibfnamefont
  {V.}~\bibnamefont {{Mandic}}}, \bibinfo {author} {\bibfnamefont
  {C.}~\bibnamefont {{Martinez}}}, \bibinfo {author} {\bibfnamefont {K.~A.}\
  \bibnamefont {{McCarthy}}}, \bibinfo {author} {\bibfnamefont
  {N.}~\bibnamefont {{Mirabolfathi}}}, \bibinfo {author} {\bibfnamefont
  {R.~A.}\ \bibnamefont {{Moffatt}}}, \bibinfo {author} {\bibfnamefont {R.~H.}\
  \bibnamefont {{Nelson}}}, \bibinfo {author} {\bibfnamefont {L.}~\bibnamefont
  {{Novak}}}, \bibinfo {author} {\bibfnamefont {K.}~\bibnamefont {{Page}}},
  \bibinfo {author} {\bibfnamefont {R.}~\bibnamefont {{Partridge}}}, \bibinfo
  {author} {\bibfnamefont {M.}~\bibnamefont {{Pepin}}}, \bibinfo {author}
  {\bibfnamefont {A.}~\bibnamefont {{Phipps}}}, \bibinfo {author}
  {\bibfnamefont {M.}~\bibnamefont {{Platt}}}, \bibinfo {author} {\bibfnamefont
  {K.}~\bibnamefont {{Prasad}}}, \bibinfo {author} {\bibfnamefont
  {M.}~\bibnamefont {{Pyle}}}, \bibinfo {author} {\bibfnamefont
  {H.}~\bibnamefont {{Qiu}}}, \bibinfo {author} {\bibfnamefont
  {W.}~\bibnamefont {{Rau}}}, \bibinfo {author} {\bibfnamefont
  {P.}~\bibnamefont {{Redl}}}, \bibinfo {author} {\bibfnamefont
  {A.}~\bibnamefont {{Reisetter}}}, \bibinfo {author} {\bibfnamefont {R.~W.}\
  \bibnamefont {{Resch}}}, \bibinfo {author} {\bibfnamefont {Y.}~\bibnamefont
  {{Ricci}}}, \bibinfo {author} {\bibfnamefont {M.}~\bibnamefont {{Ruschman}}},
  \bibinfo {author} {\bibfnamefont {T.}~\bibnamefont {{Saab}}}, \bibinfo
  {author} {\bibfnamefont {B.}~\bibnamefont {{Sadoulet}}}, \bibinfo {author}
  {\bibfnamefont {J.}~\bibnamefont {{Sander}}}, \bibinfo {author}
  {\bibfnamefont {R.~L.}\ \bibnamefont {{Schmitt}}}, \bibinfo {author}
  {\bibfnamefont {K.}~\bibnamefont {{Schneck}}}, \bibinfo {author}
  {\bibfnamefont {R.~W.}\ \bibnamefont {{Schnee}}}, \bibinfo {author}
  {\bibfnamefont {S.}~\bibnamefont {{Scorza}}}, \bibinfo {author}
  {\bibfnamefont {D.~N.}\ \bibnamefont {{Seitz}}}, \bibinfo {author}
  {\bibfnamefont {B.}~\bibnamefont {{Serfass}}}, \bibinfo {author}
  {\bibfnamefont {B.}~\bibnamefont {{Shank}}}, \bibinfo {author} {\bibfnamefont
  {D.}~\bibnamefont {{Speller}}}, \bibinfo {author} {\bibfnamefont
  {A.}~\bibnamefont {{Tomada}}}, \bibinfo {author} {\bibfnamefont
  {S.}~\bibnamefont {{Upadhyayula}}}, \bibinfo {author} {\bibfnamefont {A.~N.}\
  \bibnamefont {{Villano}}}, \bibinfo {author} {\bibfnamefont {B.}~\bibnamefont
  {{Welliver}}}, \bibinfo {author} {\bibfnamefont {D.~H.}\ \bibnamefont
  {{Wright}}}, \bibinfo {author} {\bibfnamefont {S.}~\bibnamefont {{Yellin}}},
  \bibinfo {author} {\bibfnamefont {J.~J.}\ \bibnamefont {{Yen}}}, \bibinfo
  {author} {\bibfnamefont {B.~A.}\ \bibnamefont {{Young}}}, \bibinfo {author}
  {\bibfnamefont {J.}~\bibnamefont {{Zhang}}}, \ and\ \bibinfo {author}
  {\bibnamefont {{SuperCDMS Collaboration}}},\ }\href {\doibase
  10.1103/PhysRevLett.112.241302} {\bibfield  {journal} {\bibinfo  {journal}
  {Physical Review Letters}\ }\textbf {\bibinfo {volume} {112}},\ \bibinfo
  {eid} {241302} (\bibinfo {year} {2014})},\ \Eprint
  {http://arxiv.org/abs/1402.7137} {arXiv:1402.7137 [hep-ex]} \BibitemShut
  {NoStop}%
\bibitem [{\citenamefont {{Moffat}}(2006)}]{mof06}%
  \BibitemOpen
  \bibfield  {author} {\bibinfo {author} {\bibfnamefont {J.~W.}\ \bibnamefont
  {{Moffat}}},\ }\href {\doibase 10.1088/1475-7516/2006/03/004} {\bibfield
  {journal} {\bibinfo  {journal} {J. Cosmol. Astropart. Phys.}\ }\textbf
  {\bibinfo {volume} {3}},\ \bibinfo {eid} {004} (\bibinfo {year} {2006})},\
  \Eprint {http://arxiv.org/abs/gr-qc/0506021} {gr-qc/0506021} \BibitemShut
  {NoStop}%
\bibitem [{\citenamefont {{Brownstein}}\ and\ \citenamefont
  {{Moffat}}(2006{\natexlab{a}})}]{bro+06a}%
  \BibitemOpen
  \bibfield  {author} {\bibinfo {author} {\bibfnamefont {J.~R.}\ \bibnamefont
  {{Brownstein}}}\ and\ \bibinfo {author} {\bibfnamefont {J.~W.}\ \bibnamefont
  {{Moffat}}},\ }\href {\doibase 10.1086/498208} {\bibfield  {journal}
  {\bibinfo  {journal} {The Astrophysical Journal}\ }\textbf {\bibinfo {volume}
  {636}},\ \bibinfo {pages} {721} (\bibinfo {year} {2006}{\natexlab{a}})},\
  \Eprint {http://arxiv.org/abs/astro-ph/0506370} {astro-ph/0506370}
  \BibitemShut {NoStop}%
\bibitem [{\citenamefont {{Moffat}}\ and\ \citenamefont
  {{Rahvar}}(2013)}]{mof+13}%
  \BibitemOpen
  \bibfield  {author} {\bibinfo {author} {\bibfnamefont {J.~W.}\ \bibnamefont
  {{Moffat}}}\ and\ \bibinfo {author} {\bibfnamefont {S.}~\bibnamefont
  {{Rahvar}}},\ }\href {\doibase 10.1093/mnras/stt1670} {\bibfield  {journal}
  {\bibinfo  {journal} {Mon. Not. R. Astron. Soc.}\ }\textbf {\bibinfo {volume}
  {436}},\ \bibinfo {pages} {1439} (\bibinfo {year} {2013})},\ \Eprint
  {http://arxiv.org/abs/1306.6383} {arXiv:1306.6383} \BibitemShut {NoStop}%
\bibitem [{\citenamefont {{Moffat}}\ and\ \citenamefont
  {{Toth}}(2015)}]{mof+15}%
  \BibitemOpen
  \bibfield  {author} {\bibinfo {author} {\bibfnamefont {J.~W.}\ \bibnamefont
  {{Moffat}}}\ and\ \bibinfo {author} {\bibfnamefont {V.~T.}\ \bibnamefont
  {{Toth}}},\ }\href {\doibase 10.1103/PhysRevD.91.043004} {\bibfield
  {journal} {\bibinfo  {journal} {Physical Review D}\ }\textbf {\bibinfo
  {volume} {91}},\ \bibinfo {eid} {043004} (\bibinfo {year} {2015})},\ \Eprint
  {http://arxiv.org/abs/1411.6701} {arXiv:1411.6701} \BibitemShut {NoStop}%
\bibitem [{\citenamefont {{Brownstein}}\ and\ \citenamefont
  {{Moffat}}(2006{\natexlab{b}})}]{bro+06b}%
  \BibitemOpen
  \bibfield  {author} {\bibinfo {author} {\bibfnamefont {J.~R.}\ \bibnamefont
  {{Brownstein}}}\ and\ \bibinfo {author} {\bibfnamefont {J.~W.}\ \bibnamefont
  {{Moffat}}},\ }\href {\doibase 10.1111/j.1365-2966.2006.09996.x} {\bibfield
  {journal} {\bibinfo  {journal} {Mon. Not. R. Astron. Soc.}\ }\textbf
  {\bibinfo {volume} {367}},\ \bibinfo {pages} {527} (\bibinfo {year}
  {2006}{\natexlab{b}})},\ \Eprint {http://arxiv.org/abs/astro-ph/0507222}
  {astro-ph/0507222} \BibitemShut {NoStop}%
\bibitem [{\citenamefont {{Brownstein}}\ and\ \citenamefont
  {{Moffat}}(2007)}]{bro+07}%
  \BibitemOpen
  \bibfield  {author} {\bibinfo {author} {\bibfnamefont {J.~R.}\ \bibnamefont
  {{Brownstein}}}\ and\ \bibinfo {author} {\bibfnamefont {J.~W.}\ \bibnamefont
  {{Moffat}}},\ }\href {\doibase 10.1111/j.1365-2966.2007.12275.x} {\bibfield
  {journal} {\bibinfo  {journal} {Mon. Not. R. Astron. Soc.}\ }\textbf
  {\bibinfo {volume} {382}},\ \bibinfo {pages} {29} (\bibinfo {year} {2007})},\
  \Eprint {http://arxiv.org/abs/astro-ph/0702146} {astro-ph/0702146}
  \BibitemShut {NoStop}%
\bibitem [{\citenamefont {{Moffat}}\ and\ \citenamefont
  {{Rahvar}}(2014)}]{mof+14}%
  \BibitemOpen
  \bibfield  {author} {\bibinfo {author} {\bibfnamefont {J.~W.}\ \bibnamefont
  {{Moffat}}}\ and\ \bibinfo {author} {\bibfnamefont {S.}~\bibnamefont
  {{Rahvar}}},\ }\href {\doibase 10.1093/mnras/stu855} {\bibfield  {journal}
  {\bibinfo  {journal} {Mon. Not. R. Astron. Soc.}\ }\textbf {\bibinfo {volume}
  {441}},\ \bibinfo {pages} {3724} (\bibinfo {year} {2014})},\ \Eprint
  {http://arxiv.org/abs/1309.5077} {arXiv:1309.5077} \BibitemShut {NoStop}%
\bibitem [{\citenamefont {{Moffat}}\ and\ \citenamefont
  {{Toth}}(2007)}]{mof+07}%
  \BibitemOpen
  \bibfield  {author} {\bibinfo {author} {\bibfnamefont {J.~W.}\ \bibnamefont
  {{Moffat}}}\ and\ \bibinfo {author} {\bibfnamefont {V.~T.}\ \bibnamefont
  {{Toth}}},\ }\href@noop {} {\bibfield  {journal} {\bibinfo  {journal} {ArXiv
  e-prints}\ } (\bibinfo {year} {2007})},\ \Eprint
  {http://arxiv.org/abs/0710.0364} {arXiv:0710.0364} \BibitemShut {NoStop}%
\bibitem [{Note1()}]{Note1}%
  \BibitemOpen
  \bibinfo {note} {Recently, Lopez Armengol and Romero \cite {lop+17}
  constructed neutron star models for four different equations of state, and
  were able to put restrictive upper limits on the parameter $\alpha
  $.}\BibitemShut {Stop}%
\bibitem [{\citenamefont {{Moffat}}(2015{\natexlab{a}})}]{mof15a}%
  \BibitemOpen
  \bibfield  {author} {\bibinfo {author} {\bibfnamefont {J.~W.}\ \bibnamefont
  {{Moffat}}},\ }\href {\doibase 10.1140/epjc/s10052-015-3405-x} {\bibfield
  {journal} {\bibinfo  {journal} {European Physical Journal C}\ }\textbf
  {\bibinfo {volume} {75}},\ \bibinfo {eid} {175} (\bibinfo {year}
  {2015}{\natexlab{a}})},\ \Eprint {http://arxiv.org/abs/1412.5424}
  {arXiv:1412.5424 [gr-qc]} \BibitemShut {NoStop}%
\bibitem [{\citenamefont {{Moffat}}(2015{\natexlab{b}})}]{mof15b}%
  \BibitemOpen
  \bibfield  {author} {\bibinfo {author} {\bibfnamefont {J.~W.}\ \bibnamefont
  {{Moffat}}},\ }\href {\doibase 10.1140/epjc/s10052-015-3352-6} {\bibfield
  {journal} {\bibinfo  {journal} {European Physical Journal C}\ }\textbf
  {\bibinfo {volume} {75}},\ \bibinfo {eid} {130} (\bibinfo {year}
  {2015}{\natexlab{b}})},\ \Eprint {http://arxiv.org/abs/1502.01677}
  {arXiv:1502.01677 [gr-qc]} \BibitemShut {NoStop}%
\bibitem [{\citenamefont {{Hussain}}\ and\ \citenamefont
  {{Jamil}}(2015)}]{hus+15}%
  \BibitemOpen
  \bibfield  {author} {\bibinfo {author} {\bibfnamefont {S.}~\bibnamefont
  {{Hussain}}}\ and\ \bibinfo {author} {\bibfnamefont {M.}~\bibnamefont
  {{Jamil}}},\ }\href {\doibase 10.1103/PhysRevD.92.043008} {\bibfield
  {journal} {\bibinfo  {journal} {Physical Review D}\ }\textbf {\bibinfo
  {volume} {92}},\ \bibinfo {eid} {043008} (\bibinfo {year} {2015})},\ \Eprint
  {http://arxiv.org/abs/1508.02123} {arXiv:1508.02123 [gr-qc]} \BibitemShut
  {NoStop}%
\bibitem [{\citenamefont {{John}}(2016)}]{joh16}%
  \BibitemOpen
  \bibfield  {author} {\bibinfo {author} {\bibfnamefont {A.~J.}\ \bibnamefont
  {{John}}},\ }\href@noop {} {\bibfield  {journal} {\bibinfo  {journal} {ArXiv
  e-prints}\ } (\bibinfo {year} {2016})},\ \Eprint
  {http://arxiv.org/abs/1603.09425} {arXiv:1603.09425 [gr-qc]} \BibitemShut
  {NoStop}%
\bibitem [{Note2()}]{Note2}%
  \BibitemOpen
  \bibinfo {note} {As suggested in \protect \citet {mof+13} and \cite {mof+09},
  we dismiss the scalar field $\omega $, and we treat it as a constant, $\omega
  =1$.}\BibitemShut {Stop}%
\bibitem [{Note3()}]{Note3}%
  \BibitemOpen
  \bibinfo {note} {The mass of the field $\phi ^{\mu }$ determined from galaxy
  rotation curves and galactic cluster dynamics is $\protect \mathaccentV
  {tilde}07E{\mu } = 0.042$ kpc$^{-1}$ \cite
  {mof+13,mof+14,mof+15}.}\BibitemShut {Stop}%
\bibitem [{Note4()}]{Note4}%
  \BibitemOpen
  \bibinfo {note} {Moffat \cite {mof15a} set the potential $V(\phi )$ equal to
  zero in the definition of $T^{\phi }_{\mu \nu }$ given in \cite
  {mof06}.}\BibitemShut {Stop}%
\bibitem [{Note5()}]{Note5}%
  \BibitemOpen
  \bibinfo {note} {Since the Kerr STVG metric formally resembles the
  Kerr-Newman metric in GR, the calculation of the trajectories of a test
  particle in Kerr STVG spacetime is mathematically equivalent to the
  computation of the trajectories of a test particle of electric charge $q$ in
  Kerr-Newman spacetime in GR. In this section, we adapt the formalism given in
  \cite {mis+13} to STVG.}\BibitemShut {Stop}%
\bibitem [{\citenamefont {{Lopez Armengol}}\ and\ \citenamefont
  {{Romero}}(2017)}]{lop+17}%
  \BibitemOpen
  \bibfield  {author} {\bibinfo {author} {\bibfnamefont {F.~G.}\ \bibnamefont
  {{Lopez Armengol}}}\ and\ \bibinfo {author} {\bibfnamefont {G.~E.}\
  \bibnamefont {{Romero}}},\ }\href {\doibase 10.1007/s10714-017-2184-0}
  {\bibfield  {journal} {\bibinfo  {journal} {General Relativity and
  Gravitation}\ }\textbf {\bibinfo {volume} {49}},\ \bibinfo {eid} {27}
  (\bibinfo {year} {2017})},\ \Eprint {http://arxiv.org/abs/1611.05721}
  {arXiv:1611.05721 [gr-qc]} \BibitemShut {NoStop}%
\bibitem [{\citenamefont {{Moffat}}\ and\ \citenamefont
  {{Toth}}(2008)}]{mof+08}%
  \BibitemOpen
  \bibfield  {author} {\bibinfo {author} {\bibfnamefont {J.~W.}\ \bibnamefont
  {{Moffat}}}\ and\ \bibinfo {author} {\bibfnamefont {V.~T.}\ \bibnamefont
  {{Toth}}},\ }\href {\doibase 10.1086/587926} {\bibfield  {journal} {\bibinfo
  {journal} {J. Cosmol. Astropart. Phys.}\ }\textbf {\bibinfo {volume} {680}},\
  \bibinfo {eid} {1158-1161} (\bibinfo {year} {2008})},\ \Eprint
  {http://arxiv.org/abs/0708.1935} {arXiv:0708.1935} \BibitemShut {NoStop}%
\bibitem [{\citenamefont {{Orosz}}\ \emph {et~al.}(2011)\citenamefont
  {{Orosz}}, \citenamefont {{McClintock}}, \citenamefont {{Aufdenberg}},
  \citenamefont {{Remillard}}, \citenamefont {{Reid}}, \citenamefont
  {{Narayan}},\ and\ \citenamefont {{Gou}}}]{oro+11}%
  \BibitemOpen
  \bibfield  {author} {\bibinfo {author} {\bibfnamefont {J.~A.}\ \bibnamefont
  {{Orosz}}}, \bibinfo {author} {\bibfnamefont {J.~E.}\ \bibnamefont
  {{McClintock}}}, \bibinfo {author} {\bibfnamefont {J.~P.}\ \bibnamefont
  {{Aufdenberg}}}, \bibinfo {author} {\bibfnamefont {R.~A.}\ \bibnamefont
  {{Remillard}}}, \bibinfo {author} {\bibfnamefont {M.~J.}\ \bibnamefont
  {{Reid}}}, \bibinfo {author} {\bibfnamefont {R.}~\bibnamefont {{Narayan}}}, \
  and\ \bibinfo {author} {\bibfnamefont {L.}~\bibnamefont {{Gou}}},\ }\href
  {\doibase 10.1088/0004-637X/742/2/84} {\bibfield  {journal} {\bibinfo
  {journal} {The Astrophysical Journal}\ }\textbf {\bibinfo {volume} {742}},\
  \bibinfo {eid} {84} (\bibinfo {year} {2011})},\ \Eprint
  {http://arxiv.org/abs/1106.3689} {arXiv:1106.3689 [astro-ph.HE]} \BibitemShut
  {NoStop}%
\bibitem [{\citenamefont {{Gou}}\ \emph {et~al.}(2011)\citenamefont {{Gou}},
  \citenamefont {{McClintock}}, \citenamefont {{Reid}}, \citenamefont
  {{Orosz}}, \citenamefont {{Steiner}}, \citenamefont {{Narayan}},
  \citenamefont {{Xiang}}, \citenamefont {{Remillard}}, \citenamefont
  {{Arnaud}},\ and\ \citenamefont {{Davis}}}]{gou+11}%
  \BibitemOpen
  \bibfield  {author} {\bibinfo {author} {\bibfnamefont {L.}~\bibnamefont
  {{Gou}}}, \bibinfo {author} {\bibfnamefont {J.~E.}\ \bibnamefont
  {{McClintock}}}, \bibinfo {author} {\bibfnamefont {M.~J.}\ \bibnamefont
  {{Reid}}}, \bibinfo {author} {\bibfnamefont {J.~A.}\ \bibnamefont {{Orosz}}},
  \bibinfo {author} {\bibfnamefont {J.~F.}\ \bibnamefont {{Steiner}}}, \bibinfo
  {author} {\bibfnamefont {R.}~\bibnamefont {{Narayan}}}, \bibinfo {author}
  {\bibfnamefont {J.}~\bibnamefont {{Xiang}}}, \bibinfo {author} {\bibfnamefont
  {R.~A.}\ \bibnamefont {{Remillard}}}, \bibinfo {author} {\bibfnamefont
  {K.~A.}\ \bibnamefont {{Arnaud}}}, \ and\ \bibinfo {author} {\bibfnamefont
  {S.~W.}\ \bibnamefont {{Davis}}},\ }\href {\doibase
  10.1088/0004-637X/742/2/85} {\bibfield  {journal} {\bibinfo  {journal} {The
  Astrophysical Journal}\ }\textbf {\bibinfo {volume} {742}},\ \bibinfo {eid}
  {85} (\bibinfo {year} {2011})},\ \Eprint {http://arxiv.org/abs/1106.3690}
  {arXiv:1106.3690 [astro-ph.HE]} \BibitemShut {NoStop}%
\bibitem [{\citenamefont {{Novikov}}\ and\ \citenamefont
  {{Thorne}}(1973)}]{nov+73}%
  \BibitemOpen
  \bibfield  {author} {\bibinfo {author} {\bibfnamefont {I.~D.}\ \bibnamefont
  {{Novikov}}}\ and\ \bibinfo {author} {\bibfnamefont {K.~S.}\ \bibnamefont
  {{Thorne}}},\ }in\ \href@noop {} {\emph {\bibinfo {booktitle} {Black Holes
  (Les Astres Occlus)}}},\ \bibinfo {editor} {edited by\ \bibinfo {editor}
  {\bibfnamefont {C.}~\bibnamefont {{Dewitt}}}\ and\ \bibinfo {editor}
  {\bibfnamefont {B.~S.}\ \bibnamefont {{Dewitt}}}}\ (\bibinfo {year} {1973})\
  pp.\ \bibinfo {pages} {343--450}\BibitemShut {NoStop}%
\bibitem [{\citenamefont {{Page}}\ and\ \citenamefont
  {{Thorne}}(1974)}]{pag+74}%
  \BibitemOpen
  \bibfield  {author} {\bibinfo {author} {\bibfnamefont {D.~N.}\ \bibnamefont
  {{Page}}}\ and\ \bibinfo {author} {\bibfnamefont {K.~S.}\ \bibnamefont
  {{Thorne}}},\ }\href {\doibase 10.1086/152990} {\bibfield  {journal}
  {\bibinfo  {journal} {The Astrophysical Journal}\ }\textbf {\bibinfo {volume}
  {191}},\ \bibinfo {pages} {499} (\bibinfo {year} {1974})}\BibitemShut
  {NoStop}%
\bibitem [{\citenamefont {{Misner}}\ \emph {et~al.}(1973)\citenamefont
  {{Misner}}, \citenamefont {{Thorne}},\ and\ \citenamefont
  {{Wheeler}}}]{mis+13}%
  \BibitemOpen
  \bibfield  {author} {\bibinfo {author} {\bibfnamefont {C.~W.}\ \bibnamefont
  {{Misner}}}, \bibinfo {author} {\bibfnamefont {K.~S.}\ \bibnamefont
  {{Thorne}}}, \ and\ \bibinfo {author} {\bibfnamefont {J.~A.}\ \bibnamefont
  {{Wheeler}}},\ }\href@noop {} {\emph {\bibinfo {title} {San Francisco:
  W.H.~Freeman and Co., 1973}}}\ (\bibinfo {year} {1973})\BibitemShut {NoStop}%
\bibitem [{Note6()}]{Note6}%
  \BibitemOpen
  \bibinfo {note} {We denote by $\protect \mathaccentV {tilde}07E{E}$, and
  $\protect \mathaccentV {tilde}07E{L}$ the energy and angular momentum per
  mass unit, respectively.}\BibitemShut {Stop}%
\bibitem [{\citenamefont {{Dove}}\ \emph {et~al.}(1997)\citenamefont {{Dove}},
  \citenamefont {{Wilms}}, \citenamefont {{Maisack}},\ and\ \citenamefont
  {{Begelman}}}]{dov+97}%
  \BibitemOpen
  \bibfield  {author} {\bibinfo {author} {\bibfnamefont {J.~B.}\ \bibnamefont
  {{Dove}}}, \bibinfo {author} {\bibfnamefont {J.}~\bibnamefont {{Wilms}}},
  \bibinfo {author} {\bibfnamefont {M.}~\bibnamefont {{Maisack}}}, \ and\
  \bibinfo {author} {\bibfnamefont {M.~C.}\ \bibnamefont {{Begelman}}},\ }\href
  {\doibase 10.1086/304647} {\bibfield  {journal} {\bibinfo  {journal} {The
  Astrophysical Journal}\ }\textbf {\bibinfo {volume} {487}},\ \bibinfo {pages}
  {759} (\bibinfo {year} {1997})},\ \Eprint
  {http://arxiv.org/abs/astro-ph/9705130} {astro-ph/9705130} \BibitemShut
  {NoStop}%
\bibitem [{\citenamefont {{Shakura}}\ and\ \citenamefont
  {{Sunyaev}}(1973)}]{sha+73}%
  \BibitemOpen
  \bibfield  {author} {\bibinfo {author} {\bibfnamefont {N.~I.}\ \bibnamefont
  {{Shakura}}}\ and\ \bibinfo {author} {\bibfnamefont {R.~A.}\ \bibnamefont
  {{Sunyaev}}},\ }\href@noop {} {\bibfield  {journal} {\bibinfo  {journal}
  {Astronomy and Astrophysics}\ }\textbf {\bibinfo {volume} {24}},\ \bibinfo
  {pages} {337} (\bibinfo {year} {1973})}\BibitemShut {NoStop}%
\bibitem [{\citenamefont {{P{\'e}rez}}\ \emph {et~al.}(2013)\citenamefont
  {{P{\'e}rez}}, \citenamefont {{Romero}},\ and\ \citenamefont {{Perez
  Bergliaffa}}}]{per+13}%
  \BibitemOpen
  \bibfield  {author} {\bibinfo {author} {\bibfnamefont {D.}~\bibnamefont
  {{P{\'e}rez}}}, \bibinfo {author} {\bibfnamefont {G.~E.}\ \bibnamefont
  {{Romero}}}, \ and\ \bibinfo {author} {\bibfnamefont {S.~E.}\ \bibnamefont
  {{Perez Bergliaffa}}},\ }\href {\doibase 10.1051/0004-6361/201220378}
  {\bibfield  {journal} {\bibinfo  {journal} {Astronomy and Astrophysics}\
  }\textbf {\bibinfo {volume} {551}},\ \bibinfo {eid} {A4} (\bibinfo {year}
  {2013})},\ \Eprint {http://arxiv.org/abs/1212.2640} {arXiv:1212.2640
  [astro-ph.CO]} \BibitemShut {NoStop}%
\bibitem [{\citenamefont {{Romero}}\ \emph {et~al.}(2016)\citenamefont
  {{Romero}}, \citenamefont {{Vila}},\ and\ \citenamefont
  {{P{\'e}rez}}}]{rom+16}%
  \BibitemOpen
  \bibfield  {author} {\bibinfo {author} {\bibfnamefont {G.~E.}\ \bibnamefont
  {{Romero}}}, \bibinfo {author} {\bibfnamefont {G.~S.}\ \bibnamefont
  {{Vila}}}, \ and\ \bibinfo {author} {\bibfnamefont {D.}~\bibnamefont
  {{P{\'e}rez}}},\ }\href {\doibase 10.1051/0004-6361/201527479} {\bibfield
  {journal} {\bibinfo  {journal} {Astronomy and Astrophysics}\ }\textbf
  {\bibinfo {volume} {588}},\ \bibinfo {eid} {A125} (\bibinfo {year} {2016})},\
  \Eprint {http://arxiv.org/abs/1602.07954} {arXiv:1602.07954 [astro-ph.HE]}
  \BibitemShut {NoStop}%
\bibitem [{\citenamefont {{Actis}}\ \emph {et~al.}(2011)\citenamefont
  {{Actis}}, \citenamefont {{Agnetta}}, \citenamefont {{Aharonian}},
  \citenamefont {{Akhperjanian}}, \citenamefont {{Aleksi{\'c}}}, \citenamefont
  {{Aliu}}, \citenamefont {{Allan}}, \citenamefont {{Allekotte}}, \citenamefont
  {{Antico}}, \citenamefont {{Antonelli}},\ and\ \citenamefont
  {et~al.}}]{act+11}%
  \BibitemOpen
  \bibfield  {author} {\bibinfo {author} {\bibfnamefont {M.}~\bibnamefont
  {{Actis}}}, \bibinfo {author} {\bibfnamefont {G.}~\bibnamefont {{Agnetta}}},
  \bibinfo {author} {\bibfnamefont {F.}~\bibnamefont {{Aharonian}}}, \bibinfo
  {author} {\bibfnamefont {A.}~\bibnamefont {{Akhperjanian}}}, \bibinfo
  {author} {\bibfnamefont {J.}~\bibnamefont {{Aleksi{\'c}}}}, \bibinfo {author}
  {\bibfnamefont {E.}~\bibnamefont {{Aliu}}}, \bibinfo {author} {\bibfnamefont
  {D.}~\bibnamefont {{Allan}}}, \bibinfo {author} {\bibfnamefont
  {I.}~\bibnamefont {{Allekotte}}}, \bibinfo {author} {\bibfnamefont
  {F.}~\bibnamefont {{Antico}}}, \bibinfo {author} {\bibfnamefont {L.~A.}\
  \bibnamefont {{Antonelli}}}, \ and\ \bibinfo {author} {\bibnamefont
  {et~al.}},\ }\href {\doibase 10.1007/s10686-011-9247-0} {\bibfield  {journal}
  {\bibinfo  {journal} {Experimental Astronomy}\ }\textbf {\bibinfo {volume}
  {32}},\ \bibinfo {pages} {193} (\bibinfo {year} {2011})},\ \Eprint
  {http://arxiv.org/abs/1008.3703} {arXiv:1008.3703 [astro-ph.IM]} \BibitemShut
  {NoStop}%
\bibitem [{\citenamefont {{Moffat}}\ and\ \citenamefont
  {{Toth}}(2009)}]{mof+09}%
  \BibitemOpen
  \bibfield  {author} {\bibinfo {author} {\bibfnamefont {J.~W.}\ \bibnamefont
  {{Moffat}}}\ and\ \bibinfo {author} {\bibfnamefont {V.~T.}\ \bibnamefont
  {{Toth}}},\ }\href {\doibase 10.1088/0264-9381/26/8/085002} {\bibfield
  {journal} {\bibinfo  {journal} {Classical and Quantum Gravity}\ }\textbf
  {\bibinfo {volume} {26}},\ \bibinfo {eid} {085002} (\bibinfo {year}
  {2009})},\ \Eprint {http://arxiv.org/abs/0712.1796} {arXiv:0712.1796 [gr-qc]}
  \BibitemShut {NoStop}%
\end{thebibliography}%

\end{document}